\documentclass[a4paper]{llncs}

\usepackage{amsfonts}
\usepackage{graphicx}
\usepackage{caption}
\usepackage{subcaption}
\usepackage{footnote}
\usepackage{enumerate}

\RequirePackage{graphics}
\usepackage{xcolor}
\usepackage{footnote}
\usepackage{xparse}
\usepackage{cite}
\usepackage{booktabs}

\usepackage[noend]{algpseudocode}
\usepackage{algorithm}

\algblock{ParFor}{EndParFor}
\algnewcommand\algorithmicparfor{\textbf{parfor}}
\algnewcommand\algorithmicpardo{\textbf{do}}
\algnewcommand\algorithmicendparfor{\textbf{end\ parfor}}
\algrenewtext{ParFor}[1]{\algorithmicparfor\ #1\ \algorithmicpardo}
\algrenewtext{EndParFor}{\algorithmicendparfor}
\algnewcommand\algorithmicinput{\textbf{Input:}}
\algnewcommand\algorithmicoutput{\textbf{Output:}}
\algnewcommand\Input{\item[\algorithmicinput]}%
\algnewcommand\Output{\item[\algorithmicoutput]}%

\usepackage{multicol}
\usepackage[hyphens]{url} 
\usepackage{amssymb}
\usepackage{amsmath}
\usepackage{keyval}
\usepackage{xspace}
\usepackage{paralist}
\usepackage{listings}
\usepackage{multirow}
\usepackage{stmaryrd}
\usepackage{adjustbox} 
\usepackage{makecell}
\usepackage{sidecap}
\usepackage{booktabs} 
\usepackage{threeparttable, tablefootnote}

\RequirePackage{tikz}
\usetikzlibrary{arrows,automata,shapes,calc,through,decorations.pathmorphing,decorations.fractals,chains,shapes.multipart}

\usepackage{arydshln} 
\usepackage[free-standing-units]{siunitx}
\usepackage{circuitikz}
\usepackage{tabularx} 

\def\titlename{Real-Time Verification for Distributed Cyber-Physical Systems  \xspace}
\def\authortran{Hoang-Dung Tran}
\def\authorluan{Luan Viet Nguyen}
\def\authorweiming{Weiming Xiang}

\def\authorpatrick{Patrick Musau}

\def\authortaylor{Taylor T. Johnson}

\usepackage[pdftex]{hyperref}
\hypersetup{
  colorlinks				=true,
  citecolor					={purple},
  linkcolor 				={blue},
  bookmarksnumbered	=true
}

\newcommand{\nnnum}[1]{\relax\ifmmode
  {\mathbb #1}_{\geq 0} \else ${\mathbb #1}_{\geq 0}$
  \fi}
\newcommand{\npnum}[1]{\relax\ifmmode
  {\mathbb #1}_{\leq 0} \else ${\mathbb #1}_{\leq 0}$
  \fi}
\newcommand{\pnum}[1]{\relax\ifmmode
  {\mathbb #1}_{> 0} \else ${\mathbb #1}_{> 0}$
  \fi}
\newcommand{\nnum}[1]{\relax\ifmmode
  {\mathbb #1}_{< 0} \else ${\mathbb #1}_{< 0}$
  \fi}
\newcommand{\plnum}[1]{\relax\ifmmode
  {\mathbb #1}_{+} \else ${\mathbb #1}_{+}$
  \fi}
\newcommand{\nenum}[1]{\relax\ifmmode
  {\mathbb #1}_{-} \else ${\mathbb #1}_{-}$
  \fi}

\newcommand{\extb}[1]{\relax\ifmmode {\sf ExtBeh}_{#1} \else ${\sf ExtBeh}_{#1}$\fi}
\newcommand{\tdists}[1]{\relax\ifmmode {\sf Tdists}_{#1} \else ${\sf Tdists}_{#1}$\fi}

\newcommand{\exec}[1]{\relax\ifmmode {\sf Execs}_{#1} \else ${\sf Exec}_{#1}$\fi}
\newcommand{\execf}[1]{\relax\ifmmode {\sf Execs}^*_{#1} \else ${\sf Exec}^*_{#1}$\fi}
\newcommand{\execi}[1]{\relax\ifmmode {\sf Execs}^\omega_{#1} \else ${\sf Exec}^\omega_{#1}$\fi}

\newcommand{\ctrace}[1]{\relax\ifmmode {\sf Ctraces}_{#1} \else ${\sf Ctraces}_{#1}$\fi}

\newcommand{\trace}[1]{\relax\ifmmode {\sf Traces}_{#1} \else ${\sf Traces}_{#1}$\fi}
\newcommand{\tracef}[1]{\relax\ifmmode {\sf Traces}^*_{#1} \else ${\sf Traces}^*_{#1}$\fi}
\newcommand{\tracei}[1]{\relax\ifmmode {\sf Traces}^\omega_{#1} \else ${\sf Traces}^\omega_{#1}$\fi}

\newcommand{\frag}[1]{\relax\ifmmode {\sf Frags}_{#1} \else ${\sf Frags}_{#1}$\fi}
\newcommand{\fragf}[1]{\relax\ifmmode {\sf Frags}^*_{#1} \else ${\sf Frags}^*_{#1}$\fi}
\newcommand{\fragi}[1]{\relax\ifmmode {\sf Frags}^\omega_{#1} \else ${\sf Frags}^\omega_{#1}$\fi}

\newcommand{\reach}[1]{\relax\ifmmode {\sf Reach}_{#1} \else ${\sf Reach}_{#1}$\fi}

\def\A{{\cal A}} 
\def\D{{\cal D}} 
\def\E{{\cal E}} 
\def\I{{\cal I}} 
\def\R{{\cal R}} 
\def\T{{\cal T}} 
\def\U{{\cal U}} 

\newcommand{\col}[1]{\relax\ifmmode \mathscr #1\else $\mathscr #1$\fi}

\definecolor{HIOAcolor}{rgb}{0.776,0.22,0.07}

\newcommand{\SC}[2]{\relax\ifmmode {\tt Scount}(#1,#2) \else ${\tt Scount}(#1,#2)$\fi}
\newcommand{\SCM}[2]{\relax\ifmmode {\tt Smin}(#1,#2) \else ${\tt Smin}(#1,#2)$\fi}
\newcommand{\Aut}[1]{\relax\ifmmode {\tt Aut}(#1) \else ${\tt Aut}(#1)$\fi}

\newcommand{\auto}[1]{{\operatorname{\mathsf{#1}}}}

\newcommand{\deq}{\mathrel{\stackrel{\scriptscriptstyle\Delta}{=}}}

\renewcommand{\eqref}[1]{Equation~\ref{eq:#1}}

\newcommand{\remove}[1]{}
\newcommand{\salg}[1]{\relax\ifmmode {\mathcal F}_{#1}\else ${\mathcal F}_{#1}$\fi}
\newcommand{\msp}[1]{\relax\ifmmode (#1, \salg{#1}) \else $(#1, \salg{#1})$\fi}
\newcommand{\msprod}[2]{\relax\ifmmode ( #1 \times #2, \salg{#1} \otimes \salg{#2}) \else $(#1 \times #2, \salg{#1} \otimes \salg{#2})$\fi}
\newcommand{\dist}[1]{\relax\ifmmode {\mathcal P}\msp{#1}
  \else ${\mathcal P}\msp{#1}$\fi}
\newcommand{\subdist}[1]{\relax\ifmmode {\mathcal S}{\mathcal P}\msp{#1}
  \else ${\mathcal S}{\mathcal P}\msp{#1}$\fi}
\newcommand{\disc}[1]{\relax\ifmmode {\sf Disc}(#1)
  \else ${\sf Disc}(#1)$\fi}

\newcommand{\Trajeq}{\relax\ifmmode {\mathcal R}_\T \else ${\mathcal R}_\T$\fi}
\newcommand{\Acteq}{\relax\ifmmode {\mathcal R}_A \else ${\mathcal R}_A$\fi}
\newcommand{\noop}{\relax\ifmmode \lambda \else $\lambda$\fi}
\newcommand{\close}[1]{\relax\ifmmode \overline{#1} \else $\overline{#1}$\fi}

\newcommand{\tup}[1]
           {
             \relax\ifmmode
             \langle #1 \rangle
             \else $\langle$ #1 $\rangle$ \fi
           }

\newcommand{\lit}[1]{ \relax\ifmmode
                \mathord{\mathcode`\-="702D\sf #1\mathcode`\-="2200}
                \else {\it #1} \fi }

\newcommand{\figuresize}{\scriptsize}

\lstdefinelanguage{ioa}{
  basicstyle=\figuresize,
  keywordstyle=\bf \figuresize,
  identifierstyle=\it \figuresize,
  emphstyle=\tt \figuresize,
  mathescape=true,
  tabsize=20,
  sensitive=false,
  columns=fullflexible,
  keepspaces=false,
  flexiblecolumns=true,
  basewidth=0.05em,
  escapeinside={(*@}{@*)},
  moredelim=[il][\rm]{//},
  moredelim=[is][\sf \figuresize]{!}{!},
  moredelim=[is][\bf \figuresize]{*}{*},
  keywords={automaton,and,
  	 choose,const,continue, components,
  	 discrete, do,
  	 eff, Eff, external,else, elseif, evolve, end,
  	 fi,for, forward, from,
  	 hidden,
  	 in,input,internal,if,invariant, initially, imports,
     let,
     or, output, operators, od, of,
     pre, Pre,
     return,
     such,satisfies, stop, signature, simulation,
     trajectories,trajdef, transitions, that,then, type, types, to, tasks,
     variables, vocabulary,
     when,where, with,while},
  emph={set, seq, tuple, map, array, enumeration},
   literate=
        {(}{{$($}}1
        {)}{{$)$}}1
        {\\in}{{$\in\ $}}1
        {\\preceq}{{$\preceq\ $}}1
        {\\subset}{{$\subset\ $}}1
        {\\subseteq}{{$\subseteq\ $}}1
        {\\supset}{{$\supset\ $}}1
        {\\supseteq}{{$\supseteq\ $}}1
        {\\forall}{{$\forall$}}1
        {\\le}{{$\le\ $}}1
        {\\ge}{{$\ge\ $}}1
        {\\gets}{{$\gets\ $}}1
        {\\cup}{{$\cup\ $}}1
        {\\cap}{{$\cap\ $}}1
        {\\langle}{{$\langle$}}1
        {\\rangle}{{$\rangle$}}1
        {\\exists}{{$\exists\ $}}1
        {\\bot}{{$\bot$}}1
        {\\rip}{{$\rip$}}1
        {\\emptyset}{{$\emptyset$}}1
        {\\notin}{{$\notin\ $}}1
        {\\not\\exists}{{$\not\exists\ $}}1
        {\\ne}{{$\ne\ $}}1
        {\\to}{{$\to\ $}}1
        {\\implies}{{$\implies\ $}}1
        {<}{{$<\ $}}1
        {>}{{$>\ $}}1
        {=}{{$=\ $}}1
        {~}{{$\neg\ $}}1
        {|}{{$\mid$}}1
        {'}{{$^\prime$}}1
        {\\A}{{$\forall\ $}}1
        {\\E}{{$\exists\ $}}1
        {\\nE}{{$\nexists\ $}}1
        {\\/}{{$\vee\,$}}1
        {\\vee}{{$\vee\,$}}1
        {/\\}{{$\wedge\,$}}1
        {\\wedge}{{$\wedge\,$}}1
        {=>}{{$\Rightarrow\ $}}1
        {->}{{$\rightarrow\ $}}1
        {<=}{{$\Leftarrow\ $}}1
        {<-}{{$\leftarrow\ $}}1
        {~=}{{$\neq\ $}}1
        {\\U}{{$\cup\ $}}1
        {\\I}{{$\cap\ $}}1
        {|-}{{$\vdash\ $}}1
        {-|}{{$\dashv\ $}}1
        {<<}{{$\ll\ $}}2
        {>>}{{$\gg\ $}}2
        {||}{{$\|$}}1
        {[}{{$[$}}1
        {]}{{$\,]$}}1
        {[[}{{$\langle$}}1
        {]]]}{{$]\rangle$}}1
        {]]}{{$\rangle$}}1
        {<=>}{{$\Leftrightarrow\ $}}2
        {<->}{{$\leftrightarrow\ $}}2
        {(+)}{{$\oplus\ $}}1
        {(-)}{{$\ominus\ $}}1
        {_i}{{$_{i}$}}1
        {_j}{{$_{j}$}}1
        {_{i,j}}{{$_{i,j}$}}3
        {_{j,i}}{{$_{j,i}$}}3
        {_0}{{$_0$}}1
        {_1}{{$_1$}}1
        {_2}{{$_2$}}1
        {_n}{{$_n$}}1
        {_p}{{$_p$}}1
        {_k}{{$_n$}}1
        {-}{{$\ms{-}$}}1
        {@}{{}}0
        {\\delta}{{$\delta$}}1
        {\\R}{{$\R$}}1
        {\\Rplus}{{$\Rplus$}}1
        {\\N}{{$\N$}}1
        {\\times}{{$\times\ $}}1
        {\\tau}{{$\tau$}}1
        {\\alpha}{{$\alpha$}}1
        {\\beta}{{$\beta$}}1
        {\\gamma}{{$\gamma$}}1
        {\\ell}{{$\ell\ $}}1
        {--}{{$-\ $}}1
        {\\TT}{{\hspace{1.5em}}}3
      }

\lstdefinelanguage{ioaNums}[]{ioa}
{
  numbers=left,
  numberstyle=\tiny,
  stepnumber=2,
  numbersep=4pt
}

\lstdefinelanguage{ioaNumsRight}[]{ioa}
{
  numbers=right,
  numberstyle=\tiny,
  stepnumber=2,
  numbersep=4pt
}

\lstnewenvironment{IOA}%
  {\lstset{language=IOA}}
  {}

\lstnewenvironment{IOANums}%
  {
  \if@firstcolumn
    \lstset{language=IOA, numbers=left, firstnumber=auto}
  \else
    \lstset{language=IOA, numbers=right, firstnumber=auto}
  \fi
  }
  {}

\lstnewenvironment{IOANumsRight}%
  {
    \lstset{language=IOA, numbers=right, firstnumber=auto}
  }
  {}

\newcommand{\linefigioa}[9]{

}

\lstdefinelanguage{ioaLang}{%
  basicstyle=\ttfamily\small,
  keywordstyle=\rmfamily\bfseries\small,
  identifierstyle=\small,
  keywords={assumes,automaton,axioms,backward,bounds,by,case,choose,components,const,d,det,discrete,do,eff,else,elseif,ensuring,enumeration,evolve,fi,fire,follow,for,forward,from,hidden,if,in,%
    input,initially,internal,invariant,let, local,od,of,output,pre,schedule,signature,so,%
    simulation,states,variables, tasks, stop,tasks,that,then,to,trajdef,trajectory,trajectories,transitions,tuple,type,union,urgent,uses,when,where,while,yield},
  literate=
        {\\in}{{$\in$}}1
        {\\preceq}{{$\preceq$}}1
        {\\subset}{{$\subset$}}1
        {\\subseteq}{{$\subseteq$}}1
        {\\supset}{{$\supset$}}1
        {\\supseteq}{{$\supseteq$}}1
        {\\rho}{{$\rho$}}1
        {\\infty}{{$\infty$}}1
        {<}{{$<$}}1
        {>}{{$>$}}1
        {=}{{$=$}}1
        {~}{{$\neg$}}1
        {|}{{$\mid$}}1
        {'}{{$^\prime$}}1
        {\\A}{{$\forall$}}1 {\\E}{{$\exists$}}1
        {\\/}{{$\vee$}}1 {/\\}{{$\wedge$}}1
        {=>}{{$\Rightarrow$}}1
        {->}{{$\rightarrow$}}1
        {<=}{{$\leq$}}1 {>=}{{$\geq$}}1 {~=}{{$\neq$}}1
        {\\U}{{$\cup$}}1 {\\I}{{$\cap$}}1
        {|-}{{$\vdash$}}1 {-|}{{$\dashv$}}1
        {<<}{{$\ll$}}2 {>>}{{$\gg$}}2
        {||}{{$\|$}}1
        {<=>}{{$\Leftrightarrow$}}2
        {<->}{{$\leftrightarrow$}}2
        {(+)}{{$\oplus$}}1
        {(-)}{{$\ominus$}}1
}

\lstdefinelanguage{bigIOALang}{%
  basicstyle=\ttfamily,
  keywordstyle=\rmfamily\bfseries,
  identifierstyle=,
  keywords={assumes,automaton,axioms,backward,by,case,choose,components,const,%
    d,det,discrete,do,eff,else,elseif,ensuring,enumeration,evolve,fi,for,forward,from,hidden,if,in%
    input,initially,internal,invariant,local,od,of,output,pre,schedule,signature,so,%
    tasks, simulation,states,stop,tasks,that,then,to,trajdef,trajectories,transitions,tuple,type,union,urgent,uses,when,where,yield},
  literate=
        {\\in}{{$\in$}}1
        {\\preceq}{{$\preceq$}}1
        {\\subset}{{$\subset$}}1
        {\\subseteq}{{$\subseteq$}}1
        {\\supset}{{$\supset$}}1
        {\\supseteq}{{$\supseteq$}}1
        {<}{{$<$}}1
        {>}{{$>$}}1
        {=}{{$=$}}1
        {~}{{$\neg$}}1
        {|}{{$\mid$}}1
        {'}{{$^\prime$}}1
        {\\A}{{$\forall$}}1 {\\E}{{$\exists$}}1
        {\\/}{{$\vee$}}1 {/\\}{{$\wedge$}}1
        {=>}{{$\Rightarrow$}}1
        {->}{{$\rightarrow$}}1
        {<=}{{$\leq$}}1 {>=}{{$\geq$}}1 {~=}{{$\neq$}}1
        {\\U}{{$\cup$}}1 {\\I}{{$\cap$}}1
        {|-}{{$\vdash$}}1 {-|}{{$\dashv$}}1
        {<<}{{$\ll$}}2 {>>}{{$\gg$}}2
        {||}{{$\|$}}1
        {<=>}{{$\Leftrightarrow$}}2
        {<->}{{$\leftrightarrow$}}2
        {(+)}{{$\oplus$}}1
        {(-)}{{$\ominus$}}1
}

\lstnewenvironment{BigIOA}%
  {\lstset{language=bigIOALang,basicstyle=\ttfamily}
   \csname lst@SetFirstLabel\endcsname}
  {\csname lst@SaveFirstLabel\endcsname\vspace{-4pt}\noindent}

\lstnewenvironment{SmallIOA}%
  {\lstset{language=ioaLang,basicstyle=\ttfamily\scriptsize}
   \csname lst@SetFirstLabel\endcsname}
  {\csname lst@SaveFirstLabel\endcsname\noindent}

\newlength{\bracklen}

\newcommand{\tri}[3]{\ensuremath{\mathit{#1}^\mathit{#2}_\mathit{#3}}}

\newcommand{\sugLocalVars}[2]{\ifthenelse{\equal{}{#2}}%
                             {\tri{localVars}{#1}{desug}}%
                             {\tri{localVars}{#1}{#2,desug}}}
\newcommand{\sugVars}[2]{\ifthenelse{\equal{}{#2}}%
                        {\tri{vars}{#1}{desug}}%
                        {\tri{vars}{#1}{#2,desug}}}

\newenvironment{subSyntax}{\begin{array}{l}}{\end{array}}

\newcommand{\ms}[1]{\ifmmode%
\mathord{\mathcode`-="702D\it #1\mathcode`\-="2200}\else%
$\mathord{\mathcode`-="702D\it #1\mathcode`\-="2200}$\fi}

\def\A{{\cal A}} 
\def\D{{\cal D}} 
\def\T{{\cal T}} 

\newcommand{\vv}{{\bf v}}

\newcommand{\arrow}[1]{\mathrel{\stackrel{#1}{\rightarrow}}}

\lstdefinelanguage{pvs}{
  basicstyle=\tt \figuresize,
  keywordstyle=\sc \figuresize,
  identifierstyle=\it \figuresize,
  emphstyle=\tt \figuresize,
  mathescape=true,
  tabsize=20,
  sensitive=false,
  columns=fullflexible,
  keepspaces=false,
  flexiblecolumns=true,
  basewidth=0.05em,
  moredelim=[il][\rm]{//},
  moredelim=[is][\sf \figuresize]{!}{!},
  moredelim=[is][\bf \figuresize]{*}{*},
  keywords={and,
  	 begin,
  	 cases, const,
  	 do,
  	 external, else, exists, end, endcases, endif,
  	 fi,for, forall, from,
  	 hidden,
  	 in, if, importing,
     let, lambda, lemma,
     measure,
     not,
     or, of,
     return, recursive,
     stop,
     theory, that,then, type, types, type+, to, theorem,
     var,
     with,while},
  emph={nat, setof, sequence, eq, tuple, map, array, enumeration, bool, real, exp, nnreal, posreal},
   literate=
        {(}{{$($}}1
        {)}{{$)$}}1
        {\\in}{{$\in\ $}}1
        {\\mapsto}{{$\rightarrow\ $}}1
        {\\preceq}{{$\preceq\ $}}1
        {\\subset}{{$\subset\ $}}1
        {\\subseteq}{{$\subseteq\ $}}1
        {\\supset}{{$\supset\ $}}1
        {\\supseteq}{{$\supseteq\ $}}1
        {\\forall}{{$\forall$}}1
        {\\le}{{$\le\ $}}1
        {\\ge}{{$\ge\ $}}1
        {\\gets}{{$\gets\ $}}1
        {\\cup}{{$\cup\ $}}1
        {\\cap}{{$\cap\ $}}1
        {\\langle}{{$\langle$}}1
        {\\rangle}{{$\rangle$}}1
        {\\exists}{{$\exists\ $}}1
        {\\bot}{{$\bot$}}1
        {\\rip}{{$\rip$}}1
        {\\emptyset}{{$\emptyset$}}1
        {\\notin}{{$\notin\ $}}1
        {\\not\\exists}{{$\not\exists\ $}}1
        {\\ne}{{$\ne\ $}}1
        {\\to}{{$\to\ $}}1
        {\\implies}{{$\implies\ $}}1
        {<}{{$<\ $}}1
        {>}{{$>\ $}}1
        {=}{{$=\ $}}1
        {~}{{$\neg\ $}}1
        {|}{{$\mid$}}1
        {'}{{$^\prime$}}1
        {\\A}{{$\forall\ $}}1
        {\\E}{{$\exists\ $}}1
        {\\/}{{$\vee\,$}}1
        {\\vee}{{$\vee\,$}}1
        {/\\}{{$\wedge\,$}}1
        {\\wedge}{{$\wedge\,$}}1
        {->}{{$\rightarrow\ $}}1
        {=>}{{$\Rightarrow\ $}}1
        {->}{{$\rightarrow\ $}}1
        {<=}{{$\Leftarrow\ $}}1
        {<-}{{$\leftarrow\ $}}1
        {~=}{{$\neq\ $}}1
        {\\U}{{$\cup\ $}}1
        {\\I}{{$\cap\ $}}1
        {|-}{{$\vdash\ $}}1
        {-|}{{$\dashv\ $}}1
        {<<}{{$\ll\ $}}2
        {>>}{{$\gg\ $}}2
        {||}{{$\|$}}1
        {[}{{$[$}}1
        {]}{{$\,]$}}1
        {[[}{{$\langle$}}1
        {]]]}{{$]\rangle$}}1
        {]]}{{$\rangle$}}1
        {<=>}{{$\Leftrightarrow\ $}}2
        {<->}{{$\leftrightarrow\ $}}2
        {(+)}{{$\oplus\ $}}1
        {(-)}{{$\ominus\ $}}1
        {_i}{{$_{i}$}}1
        {_j}{{$_{j}$}}1
        {_{i,j}}{{$_{i,j}$}}3
        {_{j,i}}{{$_{j,i}$}}3
        {_0}{{$_0$}}1
        {_1}{{$_1$}}1
        {_2}{{$_2$}}1
        {_n}{{$_n$}}1
        {_p}{{$_p$}}1
        {_k}{{$_n$}}1
        {-}{{$\ms{-}$}}1
        {@}{{}}0
        {\\delta}{{$\delta$}}1
        {\\R}{{$\R$}}1
        {\\Rplus}{{$\Rplus$}}1
        {\\N}{{$\N$}}1
        {\\times}{{$\times\ $}}1
        {\\tau}{{$\tau$}}1
        {\\alpha}{{$\alpha$}}1
        {\\beta}{{$\beta$}}1
        {\\gamma}{{$\gamma$}}1
        {\\ell}{{$\ell\ $}}1
        {--}{{$-\ $}}1
        {\\TT}{{\hspace{1.5em}}}3
      }

\lstdefinelanguage{BigPVS}{
  basicstyle=\tt,
  keywordstyle=\sc,
  identifierstyle=\it,
  emphstyle=\tt ,
  mathescape=true,
  tabsize=20,
  sensitive=false,
  columns=fullflexible,
  keepspaces=false,
  flexiblecolumns=true,
  basewidth=0.05em,
  moredelim=[il][\rm]{//},
  moredelim=[is][\sf \figuresize]{!}{!},
  moredelim=[is][\bf \figuresize]{*}{*},
  keywords={and,
  	 begin,
  	 cases, const,
  	 do, datatype,
  	 external, else, exists, end, endif, endcases,
  	 fi,for, forall, from,
  	 hidden,
  	 in, if, importing,
     let, lambda, lemma,
     measure,
     not,
     or, of,
     return, recursive,
     stop,
     theory, that,then, type, types, type+, to, theorem,
     var,
     with,while},
  emph={nat, setof, sequence, eq, tuple, map, array, first, rest, add, enumeration, bool, real, posreal, nnreal},
   literate=
        {(}{{$($}}1
        {)}{{$)$}}1
        {\\in}{{$\in\ $}}1
        {\\mapsto}{{$\rightarrow\ $}}1
        {\\preceq}{{$\preceq\ $}}1
        {\\subset}{{$\subset\ $}}1
        {\\subseteq}{{$\subseteq\ $}}1
        {\\supset}{{$\supset\ $}}1
        {\\supseteq}{{$\supseteq\ $}}1
        {\\forall}{{$\forall$}}1
        {\\le}{{$\le\ $}}1
        {\\ge}{{$\ge\ $}}1
        {\\gets}{{$\gets\ $}}1
        {\\cup}{{$\cup\ $}}1
        {\\cap}{{$\cap\ $}}1
        {\\langle}{{$\langle$}}1
        {\\rangle}{{$\rangle$}}1
        {\\exists}{{$\exists\ $}}1
        {\\bot}{{$\bot$}}1
        {\\rip}{{$\rip$}}1
        {\\emptyset}{{$\emptyset$}}1
        {\\notin}{{$\notin\ $}}1
        {\\not\\exists}{{$\not\exists\ $}}1
        {\\ne}{{$\ne\ $}}1
        {\\to}{{$\to\ $}}1
        {\\implies}{{$\implies\ $}}1
        {<}{{$<\ $}}1
        {>}{{$>\ $}}1
        {=}{{$=\ $}}1
        {~}{{$\neg\ $}}1
        {|}{{$\mid$}}1
        {'}{{$^\prime$}}1
        {\\A}{{$\forall\ $}}1
        {\\E}{{$\exists\ $}}1
        {\\/}{{$\vee\,$}}1
        {\\vee}{{$\vee\,$}}1
        {/\\}{{$\wedge\,$}}1
        {\\wedge}{{$\wedge\,$}}1
        {->}{{$\rightarrow\ $}}1
        {=>}{{$\Rightarrow\ $}}1
        {->}{{$\rightarrow\ $}}1
        {<=}{{$\Leftarrow\ $}}1
        {<-}{{$\leftarrow\ $}}1
        {~=}{{$\neq\ $}}1
        {\\U}{{$\cup\ $}}1
        {\\I}{{$\cap\ $}}1
        {|-}{{$\vdash\ $}}1
        {-|}{{$\dashv\ $}}1
        {<<}{{$\ll\ $}}2
        {>>}{{$\gg\ $}}2
        {||}{{$\|$}}1
        {[}{{$[$}}1
        {]}{{$\,]$}}1
        {[[}{{$\langle$}}1
        {]]]}{{$]\rangle$}}1
        {]]}{{$\rangle$}}1
        {<=>}{{$\Leftrightarrow\ $}}2
        {<->}{{$\leftrightarrow\ $}}2
        {(+)}{{$\oplus\ $}}1
        {(-)}{{$\ominus\ $}}1
        {_i}{{$_{i}$}}1
        {_j}{{$_{j}$}}1
        {_{i,j}}{{$_{i,j}$}}3
        {_{j,i}}{{$_{j,i}$}}3
        {_0}{{$_0$}}1
        {_1}{{$_1$}}1
        {_2}{{$_2$}}1
        {_n}{{$_n$}}1
        {_p}{{$_p$}}1
        {_k}{{$_n$}}1
        {-}{{$\ms{-}$}}1
        {@}{{}}0
        {\\delta}{{$\delta$}}1
        {\\R}{{$\R$}}1
        {\\Rplus}{{$\Rplus$}}1
        {\\N}{{$\N$}}1
        {\\times}{{$\times\ $}}1
        {\\tau}{{$\tau$}}1
        {\\alpha}{{$\alpha$}}1
        {\\beta}{{$\beta$}}1
        {\\gamma}{{$\gamma$}}1
        {\\ell}{{$\ell\ $}}1
        {--}{{$-\ $}}1
        {\\TT}{{\hspace{1.5em}}}3
      }

\lstdefinelanguage{pvsNums}[]{pvs}
{
  numbers=left,
  numberstyle=\tiny,
  stepnumber=2,
  numbersep=4pt
}

\lstdefinelanguage{pvsNumsRight}[]{pvs}
{
  numbers=right,
  numberstyle=\tiny,
  stepnumber=2,
  numbersep=4pt
}

\lstnewenvironment{BigPVS}%
  {\lstset{language=BigPVS}}
  {}

\lstnewenvironment{PVSNums}%
  {
  \if@firstcolumn
    \lstset{language=pvs, numbers=left, firstnumber=auto}
  \else
    \lstset{language=pvs, numbers=right, firstnumber=auto}
  \fi
  }
  {}

\lstnewenvironment{PVSNumsRight}%
  {
    \lstset{language=pvs, numbers=right, firstnumber=auto}
  }
  {}

\newcommand{\linefigpvs}[9]{

}

\lstdefinelanguage{pvsproof}{
  basicstyle=\tt \figuresize,
  mathescape=true,
  tabsize=4,
  sensitive=false,
  columns=fullflexible,
  keepspaces=false,
  flexiblecolumns=true,
  basewidth=0.05em,
}

\def\N{\act{N}}

\newcommand{\localvar}[2]{{{#1_{#2}}}}

\def\xi{\localvar{x}{i}}

\def\reach{{\sf Reach}}

\newcommand{\valvi}{val{(V_i)}}

\def\Xi{\mathit{X_i}}

\newcommand{\clk}{{\mathit clk}}
\newcommand{\msgs}{{\mathit msghist}}

\begin{document}
%

\title{\titlename}

%
%
%

\author{\authortran \inst{1}, \authorluan \inst{2}, \authorpatrick \inst{1}, \authorweiming \inst{3}, \and \authortaylor \inst{1}}

\authorrunning{Tran et al.}
%
\institute{Institute for Software Integrated Systems, Vanderbilt University, TN, USA \and
  Department of Computer Science and Engineering, University of Notre Dame, USA \and
  School of Computer and Cyber Sciences, Augusta University, USA}


%
%
\maketitle

\begin{abstract}
\vspace{-1em}
	Safety-critical distributed cyber-physical systems (CPSs) have been found in a wide range of applications. Notably, they have displayed a great deal of utility in intelligent transportation, where autonomous vehicles communicate and cooperate with each other via a high-speed communication network. Such systems require an ability to identify maneuvers in real-time that cause dangerous circumstances and ensure the implementation always meets safety-critical requirements. In this paper, we propose a real-time decentralized reachability approach for safety verification of a distributed multi-agent CPS with the underlying assumption that all agents are time-synchronized with a low degree of error. In the proposed approach, each agent periodically computes its local reachable set and exchanges this reachable set with the other agents with the goal of verifying the system safety. Our method,  implemented in Java, takes advantages of the timing information and the reachable set information that are available in the exchanged messages to reason about the safety of the whole system in a decentralized manner. Any particular agent can also perform local safety verification tasks based on their local clocks by analyzing the messages it receives. We applied the proposed method to verify, in real-time, the safety properties of a group of quadcopters performing a distributed search mission.

\end{abstract}

\section{Introduction}
\label{intro}

The emergence of 5G technology has inspired a massive wave of the research and development in science and technology in the era of IoT where the communication between computing devices has become significantly faster with lower latency and power consumption. The power of this modern communication technology influences and benefits all aspects of Cyber-Physical Systems (CPSs) such as smart grids, smart homes, intelligent transportation and smart cities. In particular, the study of autonomous vehicles has become an increasingly popular research field in both academic and industrial transportation applications. Automotive crashes pose significant financial and life-threatening risks, and there is an urgent need for advanced and scalable methods that can efficiently verify a distributed system of autonomous vehicles.

Over the last two decades, although many methods have been developed to conduct reachability analysis and safety verification of CPS, such as the approaches proposed in~\cite{le2009reachability, girard2006efficient, althoff2015introduction, henzinger1997hytech,frehse2011spaceex, chen2013flow, kong2015dreach, bak2017hylaa, bak2017simulation, tran2017order, tran2019formats}, applying these techniques to \emph{real-time distributed} CPS remains a big challenge. This is due to the fact that, 1) all existing techniques have intensive computation costs and are usually too slow to be used in a real-time manner and, 2) these techniques target the safety verification of a \emph{single} CPS, and therefore they naturally cannot be applied efficiently to a \emph{distributed} CPS where clock mismatches and communication between agents (i.e., individual systems) are essential concerns. Since the future autonomous vehicles systems will work distributively involving effective communication between each agent, there is an urgent need for an approach that can provide formal guarantees of the safety of distributed CPS in real-time. More importantly, the safety information should be defined based on the \emph{agents local clocks} to allow these agents to perform ``intelligent actions'' to escape from the upcoming dangerous circumstances. For example, if an agent A knows based on its local clock that it will collide with an agent B in the next 5 seconds, it should perform an action such as stopping or quickly finding a safe path to avoid the collision.

In this paper\footnote{This paper is an extension of \cite{tran2019forte}}, we propose a \emph{decentralized real-time reachability} approach for safety verification of a distributed CPS with multiple agents. We are particularly interested in two types of safety properties. The first one is a \emph{local safety property} which specifies the local constraints of the agent operation. For example, each agent is only allowed to move within a specific region, does not hit any obstacles, and its velocity needs to be limited to specific range. This type of property does not require the information of other agents and can be verified locally at run-time. The second safety property is a \emph{global property} defined on the states of multiple agents. Particularly, we consider a peer-to-peer collision free property and a generalized property where we want to verify if all agents satisfy a set of linear constraints (on the states of all agents) defining the property, e.g., two agents do not go into the same region at the same time.

Our decentralized real-time reachability approach works as follows. Each agent \emph{locally} and \emph{periodically computes} the local reachable set from the current local time to the next $T$ seconds, and then \emph{encodes} and \emph{broadcasts} its reachable set information to the others via a communication network. When the agent receives a reachable set message, it immediately \emph{decodes} the message to read the reachable set information of the sender, and then performs \emph{peer-to-peer collision checking} based on its current state and the reachable set of the sender. Verifying a generalized global property involving the states of $N$ agents is done at the time an agent receives all needed reachable sets from other agents. Additionally, the local safety property of the agent is verified simultaneously with the reachable set computation process at run-time. The proposed verification approach is based on an underlying assumption that is, all agents are time-synchronized to some level of accuracy. This assumption is reasonable as it can be achieved by using existing time synchronization protocols such as the Network Time Protocol (NTP). Our approach has successfully verified in real-time the local safety properties and collision occurrences for a group of quadcopters conducting a search mission.

The rest of the paper is organized as follows. Section \ref{sec:modeling} presents briefly the distributed CPS modeling and its verification problems. Section \ref{sec:face-lifting} gives the detail of real-time reachability for single agent and how to use it for real-time local safety verification. Section \ref{sec:collision-checking} addresses the utilization reachable set messages for checking peer-to-peer collision. Section \ref{sec:global-verification} investigates the global safety verification problem. Section \ref{sec:evaluation} presents the implementation and evaluation of our approach via a distributed search application using quadcopters.
%

%

\section{Problem Formulation}
\label{sec:modeling}
In this paper, we consider a distributed CPS with $N$ agents that can communicate with each other via an asynchronous communication channel. 


\paragraph*{Communication Model}
The communication between agents is implemented by the \emph{actions} of sending and receiving messages over an asynchronous communication channel. We formally model this communication model as a single automaton, $\auto{Channel}$, which stores the set of in-flight messages that have been sent, but are yet to be delivered. When an agent sends a message $m$, it invokes a \emph{send(m)} action. This action adds $m$ to the \emph{in-flight} set. At any arbitrary time, the $\auto{Channel}$ chooses a message in the in-flight set to either delivers it to its recipient or removes it from the set. All messages are assumed to be unique and each message contains its sender and recipient identities. Let $M$ be the set of all possible messages used in communication between agents. The sending and receiving messages by agent $i$ are denoted by $M_{i,*}$ and $M_{*,i}$, respectively.

\paragraph*{Agent Model}
The $i^{th}$ agent is modeled as a hybrid automaton~\cite{henzinger1996lics, lynch1996hybrid} defined by the tuple $\langle\A_i = V_i, A_i, \D_i, \T_i \rangle$, where:
\begin{enumerate}[a)]
\item $V_i$ is a set of variables consisting of the following:
	\begin{inparaenum}[i)]
		\item a set of continuous variables $X_i$ including a special variable $\clk_i$ which records the agent's \emph{local time}, and
		\item a set of discrete variables $Y_i$ including the special variable $\msgs_i$ that records all sent and received messages. A valuation $\vv_i$ is a function that associates each $v_i \in V_i$ to a value in its type. We write $\valvi$ for the set of all possible valuations of $V_i$. We abuse the notion of $\vv_i$ to denote a state of $\A_i$, which is a valuation of all variables in $V_i$.
	\end{inparaenum}
	The set  $Q_i \deq val(V_i)$ is called the set of \emph{states}.
\item $A_i$ is a set of \emph{actions} consisting of the following subsets:
	\begin{inparaenum}[i)]
		\item a set $\{ \emph{send}_i(m) \ | \ m \in M_{i,*} \}$ of send actions (i.e., output actions),
		\item a set $\{ \emph{receive}_i(m) \ | \ m \in M_{*,i} \}$ of receive actions (i.e., input actions), and
		\item a set $H_i$ of other, ordinary actions.
		\end{inparaenum}
\item $\D_i \subseteq val(V_i) \times A_i \times val(V_i)$ is called the set of {\em transitions\/}.
For a transition $(\vv_i, a_i, \vv_i') \in \D_i$, we write $\vv_i \arrow{a_i} \vv_i'$ in short.
		\begin{inparaenum}[i)]
		\item If $a_i = \emph{send}_i(m)$ or $\emph{receive}_i(m)$, then all the components of $\vv_i$ and $\vv_i'$ are identical except that $m$ is added to $\msgs$ in $\vv_i'$. That is, the agent's other states remain the same on message sends and receives.
		Furthermore, for every state $\vv_i$ and every receive action $a_i$, there must exist a $\vv_i'$ such that $\vv_i \arrow{a_i} \vv_i'$, i.e., the automaton must have well-defined behavior for receiving any message in any state.
		\item If $a_i \in H_i$, then $\vv_i.\msgs = \vv_i'.\msgs$.
		\end{inparaenum}
\item $\T_i$ is a collection of trajectories for $X_i$. Each trajectory of $X_i$ is a function mapping an interval of time $[0,t], t \geq 0$ to $\valvi$, following a flow rate that specifies how a real variable $x_i \in X_i$ evolving over time. We denote the \emph{duration} of a trajectory as $\tau_{dur}$, which is the right end-point of the interval $t$.
\end{enumerate}

\paragraph*{Agent Semantics}
The \emph{behavior} of each agent can be defined based on the concept of an \emph{execution} which is a particular run of the agent. Given an initial state $\vv^0_i$, an \emph{execution} $\alpha_i$ of an agent $A_i$ is a sequence of states starting from $\vv^0_i$, defined as $\alpha_i = \vv^0_i, \vv^1_i, \ldots$, and for each index $j$ in the sequence, the state update from $\vv^j_i$ to $\vv^{j+1}_i$ is either a transition or trajectory. A state $\vv^j_i$ is \emph{reachable} if there exists an executing that ends in $\vv^j_i$. We denote $\reach(A_i)$ as the reachable set of agent $A_i$.

\paragraph*{System Model}
The formal model of the complete system, denoted as $\auto{System}$, is a network of hybrid automata that is obtained by parallel composing the agent's models and the communication channel. Formally, we can write, $\auto{System} \deq \A_1 \| \ldots \A_N \| \auto{Channel}$. Informally, the agent $\A_i$ and the communication channel $\auto{Channel}$ are synchronized through sending and receiving actions. When the agent $A_i$ sends a message $m \in M_{i,j}$ to the agent $A_j$, it triggers the $\emph{send}_i(m)$ action. At the same time, this action is synchronized in the  $\auto{Channel}$ automaton by putting the message $m$ in the \emph{in-flight} set. After that, the $\auto{Channel}$ will trigger (non-deterministically) the $\emph{receive}_j(m)$ action. This action is synchronized in the agent $A_j$ by putting the message $m$ into the $\msgs_j$. 
%

In this paper, we investigate three real-time safety verification problems for distributed cyber-physical systems as defined in the following.

\begin{problem}[Local safety verification in real-time]
The real-time local safety verification problem is to compute online the reachable set $\reach(A_i)$ of the agent and verify if it violates the local safety property, i.e., checking $\reach(A_i) \cap \mathcal{U}_i = \emptyset ?$, where $\mathcal{U}_i \triangleq C_ix_i \leq d_i, x_i \in X_i$ is the unsafe set of the agent.
\end{problem}
\begin{problem}[Decentralized real-time collision verification]
The decentralized real-time collision verification problem is to reason in real-time whether an agent $A_i$ will collide with other agents from its current local time $t_c^i$ to the \emph{computable, safe time instance in the future} $T_{safe}$ based on
\begin{inparaenum}[i)]
	\item the \emph{clock mismatches}, and
	\item the \emph{exchanging reachable set messages} between agents.
\end{inparaenum}
Formally, we require that $\forall~ t_c^i \leq t \leq T_{safe}, d_{ij}(t) \geq l$, where $d_{ij}(t)$ is the distance between agents $A_i$ and $A_j$ at the time $t$ of the agent $A_i$ local clock, and $l$ is the allowable safe distance between agents.
\end{problem}

\begin{problem}[Decentralized real-time global safety verification]
The decentralized real-time global safety verification problem is to construct online (at each agent) the reachable set of all agents $globalReach$ and verify if it violates the global safety property, i.e., checking $globalReach \cap \mathcal{U} = \emptyset$, where $\mathcal{U} \triangleq Cx \leq d$, $x = [x_1^T, \dots, x^T_N]^T, x_i \in X_i$, is the unsafe set of the whole system.

\end{problem}


\section{Real-Time Local Safety Verification}
\label{sec:face-lifting}

The first important step in our approach is, each agent $A_i$ computes forwardly its reachable set of states from the current local time $t^i$ to the next $(t^i + T)$ seconds which is defined by $\mathcal{R}_i[t^i, t^i + T]$. Since there are many variables used in the agent modeling that are irrelevant in safety verification, we only need to compute the reachable set of state that is related to the agent's physical dynamics (so called as \emph{motion dynamics}) which is defined by a nonlinear ODE $\dot{x}_i = f(x_i, u_i)$, where $x_i \in \mathbb{R}^n$ is state vector and $u_i \in \mathbb{R}^m$ is the control input vector. The agent can switch from one mode to the another mode via discrete transitions, and in each mode, the control law may be different. When the agent computes its reachable set, the only information it needs are its current set of states $x_i(t^i)$ and the current control input $u_i(t^i)$. It should be clarified that although the control law may be different among modes, the control signal $u_i$ is updated with the same control period $T^i_c$. Consequently, $u_i$ is a constant vector in each control period.

Assuming that the agent's current time is $t^i_j = j \times T_c$, using its local sensors and GPS, we have the current state of the agent $x_i$. Note that the local sensors and the provided GPS can only provide the information of interest to some accuracy, therefore the actual state of the agent is in a set $x_i \in I_i$. The control signal $u_i$ is computed based on the state $x_i$ and a reference signal, e.g., a set point denoting where the agent needs to go to, and then computed control signal is applied to the actuator to control the motion of the agent. From the current set of states $I_i$ and the control signal $u_i$, we can compute the forward reachable set of the agent for the next $t^i_j + T$ seconds. This reachable set computation needs to be completed after an amount of time $T^i_{runtime} < T^i_c$ because if $T^i_{runtime} \geq T^i_c$, a new $u_i$ will be updated. The control period $T^i_c$ is chosen based on the agent's motion dynamics, and thus to control an agent with fast dynamics, the control period $T^i_c$ needs to be sufficiently small. This is the source of the requirement that the allowable run-time for reachable set computation be small.

To compute the reachable set of an agent in real-time, we use the well-known face-lifting method~\cite{dang1998hscc, bak2014real} and a \emph{hyper-rectangle} to represent the reachable set. This method is useful for short-time reachability analysis of real-time systems. It allows users to define an allowable run-time $T^i_{runtime}$, and has no dynamic data structures, recursion, and does not depend on complex external libraries as in other reachability analysis methods. More importantly, the accuracy of the reachable set computation can be iteratively improved based on the \emph{remaining allowable run-time}.
\begin{algorithm}[t]\small
    \caption{Real-time reachability analysis for agent $A_i$.}
    \label{arg:real-time-reach}
		\begin{flushleft}
				\textbf{Input}: $I_i$, $u_i$, $t^i$, $T$, $h_i$, $T^i_{runtime}$, $\mathcal{U}_i$\\
				\textbf{Output}: $\mathcal{R}_i[t^i, t^i + T]$, $safe = true$ or $safe = uncertain$
		\end{flushleft}
    \begin{algorithmic}[1]
        \Procedure{Initialization}{}
        \State    $step = h_i$ ~~~~~~~~~\% {Reach time step}
        \State    $T^i_1 = T^i_{runtime}$ ~~~\% {Remaining run-time}
        \EndProcedure

        \Procedure{Reachability Analysis}{}
        \While{$(T^i_1 > 0)$}
            \State $\mathcal{CR} = I_i$ ~~~~~~~\% {Current reachable set}
            \State $safe = true$
            \State $T^i_2 = T$ ~~~~~~~~\% {Remaining reach time}
            \While{$T^i_2 > 0$}
                \State \% {Do Single Face Lifting}
                \State $\mathcal{R}, T' = SFL(\mathcal{CR}, step , T^i_2, u_i)$
                \State $\mathcal{CR} = \mathcal{R}$ ~~~\% {Update reach set}
                \State $T^i_2 = T'$ ~~~\% {Update remaining reach time}
                \If{$(\mathcal{CR} \cap \mathcal{U}_i \neq \emptyset)$}: $safe = uncertain$
                \EndIf
                \State $\mathcal{R}_i[t^i, t^i + T] = \mathcal{CR}$
            \EndWhile
            \State \% \textit{Update remaining runtime}
            \State $T_1^i = T_1^i - (A_i.currentTime() - t^i)$
            \If{$T_1^i \leq 0$} break
            \Else \State $step = h_i / 2$ ~~~\% {Reduce reach time step}
            \EndIf

        \EndWhile
        \State \textbf{return} $\mathcal{R}_i[t^i, t^i + T] = \mathcal{CR}, safe$
    \EndProcedure
    \end{algorithmic}
\end{algorithm}

Algorithm \ref{arg:real-time-reach} describes the real-time reachability analysis for one agent. The Algorithm works as follows. The time period $[t^i, t^i + T]$ is divided by $M$ steps. The reach time step is defined by $h_i = T/M$. Using the reach time step and the current set $I_i$, the face-lifting method performs a \emph{single-face-lifting operation}. The results of this step are a new reachable set and a \emph{remaining reach time} $T^i_{remainReachTime} < T$. This step is iteratively called until the reachable set for the whole time period of interest $[t^i, t^i + T]$ is constructed completely, i.e., the remaining reach time is equal to zero. Interestingly, with the reach time step size $h_i$ defined above, the face-lifting algorithm may be finished quickly after an amount of time which is smaller than the allowable run-time $T^i_{runtime}$ specified by user, i.e., there is still an amount of time called remaining run time $T^i_{remainRunTime} < T^i_{runtime}$ that is available for us to recall the face-lifting algorithm with a smaller reach time step size, for example, we can recall the face-lifting algorithm with a new reach time step $h_i / 2$. By doing this, the conservativeness of the reachable set can be iteratively improved. The core step of face-lifting method is the single-face-lifting operation. We refer the readers to~\cite{bak2014real} for further detail. As mentioned earlier, the local safety property of each agent can be verified at run-time simultaneously with the reachable set computation process. Precisely, let $\mathcal{U}_i \triangleq C_ix_i \leq d_i$ be the unsafe region of the $i^{th}$ agent, the agent is said to be safe from $t^i$ to $t^i + t \leq t^i + T$ if $\mathcal{R}_i[t^i, t^i + t] \cap \mathcal{U}_i = \emptyset$. Since the reachable set $\mathcal{R}_i[t^i, t^i + t]$ is given by the face-lifting method at run-time, the local safety verification problem for each agent can be solved at run-time. Since the Algorithm \ref{arg:real-time-reach} computes an over-approximation of the reachable set of each agent in a short time interval, it guarantees the soundness of the result as described in the following lemma.

\begin{lemma}\cite{dang1998hscc, bak2014real}\label{lm:individual-soundness}
The real-time reachability analysis algorithm is sound, i.e., the computed reachable set contains all possible trajectories of agent $A_i$ from $t^i$ to $t^i + T$.
\end{lemma}

%


\section{Decentralized Real-Time Collision Verification}
\label{sec:collision-checking}

\begin{figure}[t]
	\centering
		\includegraphics[width=0.8\textwidth]{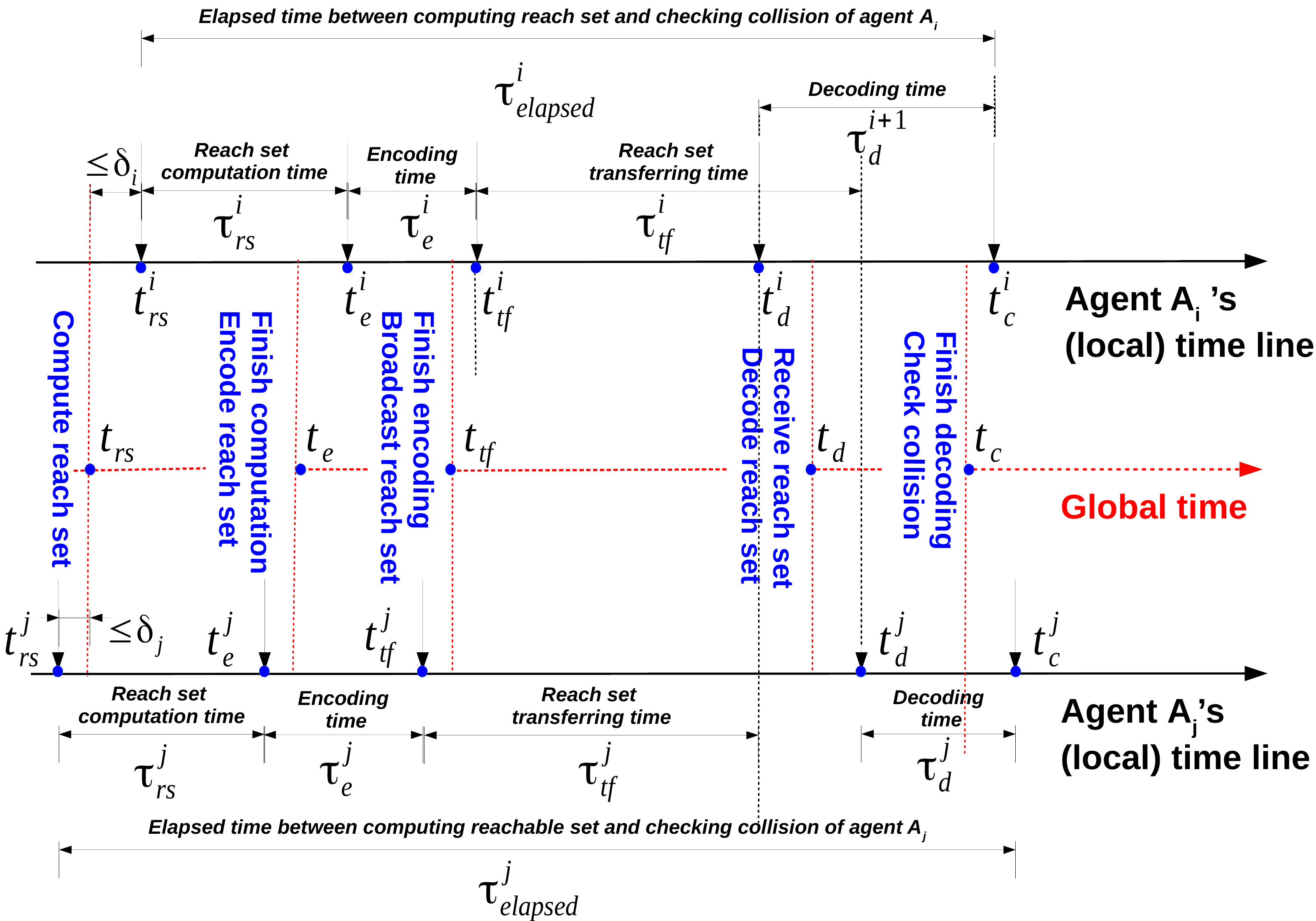}
		\caption{Timeline for reachable set computing, encoding, transferring, decoding and collision checking.}
        \label{fig:drreach-events}
         \vspace{-1em}
\end{figure}

Our collision verification scheme is performed based on the exchanged reachable set messages between agents. For every control period $T_c$, each agent executes the real-time reachability analysis algorithm to check if it is locally safe and to obtain its current reachable set with respect to its current control input. When the current reachable set is available, the agent encodes the reachable set in a message and then broadcasts this message to its cooperative agents and listens to the upcoming messages sent from these agents. When a reachable set message arrives, the agent immediately decodes the message to construct the current reachable set of the sender and then performs peer-to-peer collision detection. The process of computing, encoding, transferring, decoding of the reachable set along with collision checking is illustrated in Figure~\ref{fig:drreach-events} based on the agent's local clock.

Let $t^i_{rs}$, $t^i_e$, $t^i_{tf}$, $t^i_d$, and $t^i_c$ respectively be the instants that we compute, encode, transfer, decode the reachable set and do collision checking on the agent $A_i$. Note that these time instants are based on the agent $A_i$'s local clock. The actual run-times are defined as follows.
\begin{equation*}
\begin{split}
&\tau^i_{rs} = t^i_e - t^i_{rs}, \%~\textit{reachablet set computation~time}, \\
&\tau^i_e = t^i_{tf} - t^i_{e}, \%~\textit{encoding time}, \\
&\tau^i_{tf} \approx t^j_d - t^i_{tf}, \%~\textit{transferring time}, \\
&\tau^i_{d} = t^i_c - t^i_d, \%~\textit{decoding time}.
\end{split}
\end{equation*}

Note that we do not know the exact transfer time $\tau^i_{tf}$ since it depends on two different local time clocks. The above transfer time formula describes its approximate value when neglecting the mismatch between the two local clocks. The actual reachable set computation time is close to the allowable run-time chosen by user, i.e., $\tau^i_{rs} \approx T^i_{runtime}$. We will see later that the encoding time and decoding time are fairly small in comparison with the transferring time, i.e., $\tau^i_e \approx \tau^i_d \ll \tau^i_{tf}$. All of these run-times provide useful information for selecting an appropriate control period $T_c$ for an agent. However, for collision checking purpose, we only need to consider the time instants that an agent starts computing reachable set $t^i_{rs}$ and checking collision $t^i_c$.

A reachable set message contains three pieces of information: the reachable set which is a list of intervals, the time period (based on the local clock) in which this reachable set is valid, i.e., the start time $t^i_{rs}$ and the end time $t^i_{rs} + T$ and the time instant that this message is sent. Based on the timing information of the reachable set and the time-synchronization errors, an agent can examine whether or not a received reachable set contains information about the future behavior of the sent agent which is useful for checking collision. The usefulness of the reachable sets used in collision checking is defined as follows.
\begin{figure}[t]
	\centering
		\includegraphics[width=0.8\textwidth]{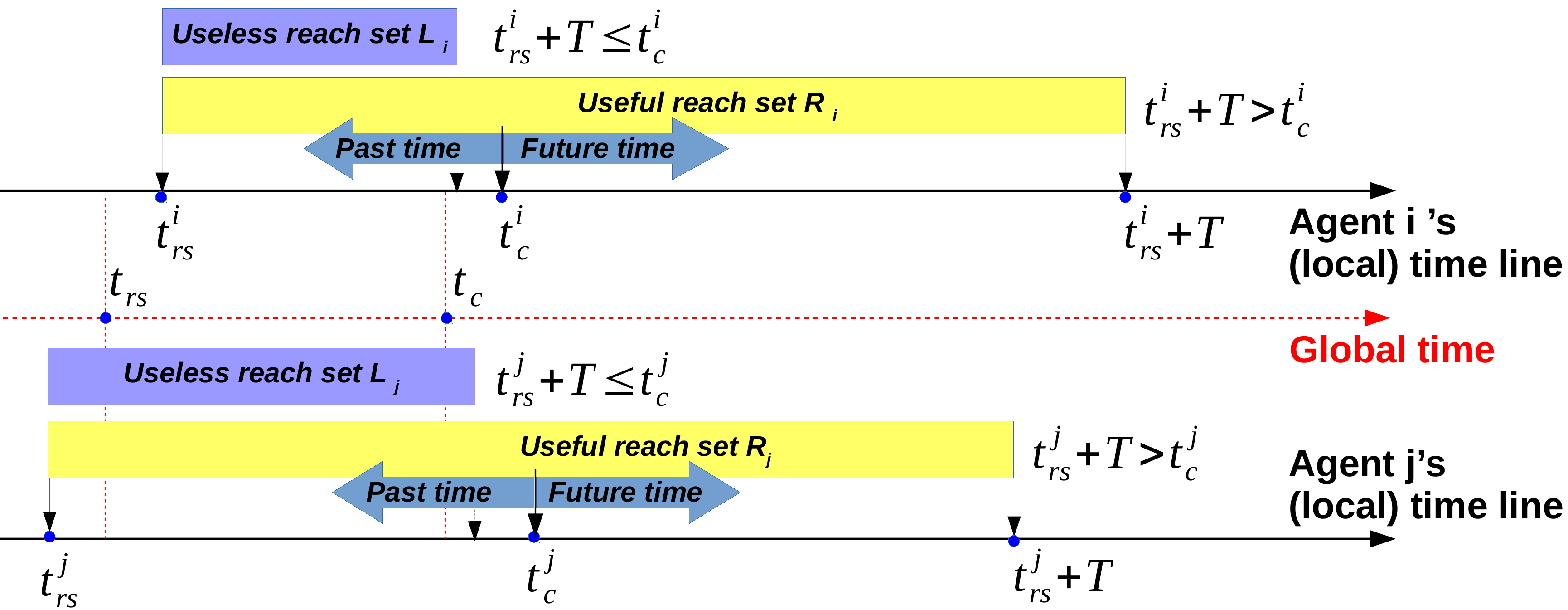}
		\caption{Useful reachable set.}
        \label{fig:useful-reach-set}
         \vspace{-1em}
 \end{figure}

\begin{definition}[Useful reachable sets]
Let $\delta_i$ and $\delta_j$ respectively be the time-synchronization errors of agent $A_i$ and $A_j$ in comparison with the \emph{virtual global time t}, i.e, $t - \delta_i \leq t^i \leq t + \delta_i$ and $t - \delta_j \leq t^j \leq t + \delta_j$, where $t^i$ and $t^j$ are current local times of $A_i$ and $A_j$ respectively. The reachable sets $\mathcal{R}_i[t^i_{rs}, t^i_{rs} + T]$ and $\mathcal{R}_j[t^j_{rs}, t^j_{rs} + T]$ of the agent $A_j$ that are available at the agent $A_i$ at time $t^i_c$ are \emph{useful} for checking collision between $A_i$ and $A_j$ if:
\begin{equation}\label{useful-condition}
\begin{split}
&t^i_c < t^j_{rs} + T - \delta_i - \delta_j, \\
&t^i_c < t^i_{rs} + T.
\end{split}
\end{equation}
\end{definition}
%
Assume that we are at a time instant where the agent $A_i$ checks if a collision occurs. This means that the current local time is $t^i_c$. Note that agent $A_i$ and $A_j$ are synchronized to the global time with errors $\delta_i$ and $\delta_j$ respectively. The reachable set  $\mathcal{R}_j[t^j_{rs}, t^j_{rs} + T]$ is useful if it contains information about the \emph{future behavior} of agent $A_j$ under the view of the agent $A_i$ based on its local clock. This can be guaranteed if we have:  $t^j_{rs} + T \geq t^{i}_{rs} - \delta_j + T > t^i_c + \delta_i$. Additionally, the current reachablet set of agent $A_i$ contains information about its future behavior if $t^i_c < t^i_{rs} + T$ as depicted in Figure \ref{fig:useful-reach-set}. We can see that if $t^i_c > t^j_{rs} + T + \delta_i + \delta_j$, then the reachable set of $A_j$ contains a past information, and thus it is useless for checking collision. One interesting case is when $ t^j_{rs} + T - \delta_i - \delta_j < t^i_c < t^j_{rs} + T + \delta_i + \delta_j$. In this case, we do not know whether the received reachable set is useful or not.

\begin{remark} We note that the proposed approach does not rely on the concept of Lamport happens-before relation~\cite{lamport1978time} to compute the local reachable set of each agent. If the agent could not receive reachable messages from others until a requested time-stamp expires, it still calculates the local reachable set based on its current state and the state information of other agents in the messages it received previously. In other words, our method does not require the reachable set of each agent to be computed corresponding to the ordering of the events (sending or receiving a message) in the system, but only relies on the local clock period and the time-synchronization errors between agents. Such implementation ensures that the computation process can be accomplished in real-time, and is not affected by the message transmission delay.
\end{remark}
\begin{algorithm}[t]\small
    \caption{Decentralized Real-Time Collision Verification at Agent $A_i$.}
    \label{arg:collision-detection}
		\begin{flushleft}
    \textbf{Input}:  $l$,~~\% safe distance between agents \\
    \textbf{Output}: $collision, T_{safe}$ ~~\% collision flag and safe time interval in the future
		\end{flushleft}
    \begin{algorithmic}[1]

        \Procedure{Peer-to-Peer Collision Detection}{}
            \If{new message $\mathcal{R}_j[t^j_{rs}, t^j_{rs} + T]$ arrive}
            \State decode message
            \State $t^i_c = A_i.current\_time()$ ~~\% current time
            \State $t^i_{rs} = \mathcal{R}_i.t^i_{rs}$ ~~\% current reachable set start time
            \If{$t^i_c < t^j_{rs} + T - \delta_i - \delta_j$ and $t^i_c < t^i_{rs} + T$} ~~\% check usefulness
            \State compute possible minimum distance $d_{min}$ between two agents
            \If{$d_{min} > l$}
            \State Collision = false
            \State  $T_{safe} = min(t^j_{rs} + T - \delta_i - \delta_j, t^i_{rs} + T)$
            \Else
            \State Collision = uncertain, $T_{safe} = [~]$
            \EndIf
            \State store the message
            \EndIf
            \EndIf
        \EndProcedure

    \end{algorithmic}
\end{algorithm}

The peer-to-peer collision checking procedure depicted in Algorithm \ref{arg:collision-detection} works as follows: when a new reachable set message arrives, the receiving agent decodes the message and checks the usefulness of the received reachable set and its current reachable set. Then, the agent combines its current reachable set and the received reachable set to compute the minimum possible distance between two agents. If the distance is larger than an allowable threshold $l$, there is no collision between two agents in some known time interval in the future, i.e., $T_{safe}$.

\begin{lemma}\label{lm:collision}
  The decentralized real-time collision verification algorithm is sound.
\begin{proof}
From Lemma \ref{lm:individual-soundness}, we know that the received reachable set $\mathcal{R}_j[t^j_{rs}, t^j_{rs} + T]$ contains all possible trajectories of the agent $A_j$ from $t^j_{rs}$ to $t^j_{rs} + T$. Also, the current reachable set of the agent $A_i$, $\mathcal{R}_i[t^i_{rs}, t^i_{rs} + T]$, contains all possible trajectories of the agent from $t^i_{rs}$ to $t^i_{rs} + T$. If those reachable sets are useful, then they contains all possible trajectories of two agents from $t_c^i$ to sometime $T_{safe} = min(t^j_{rs} + T - \delta_i - \delta_j, t^i_{rs} + T)$ in the future based on the agent $A_i$ clock. Therefore, the minimum distance $d_{min}$ between two agents computed from two reachable sets is the smallest distance among all possible distances in the time interval $[t^i_{c},T_{safe}]$. Consequently, the collision free guarantee is sound in the time interval $[t^i_{c},T_{safe}]$.
\end{proof}
\end{lemma}

We have studied how to use exchanged reachable sets to do peer-to-peer collision detection. Next, we consider how to verify online the global behavior of a distributed CPS in decentralized manner.

\section{Decentralized Real-Time Global Safety Verification}
\label{sec:global-verification}
\begin{definition}[Globally useful reachable set.] Consider a distributed CPS with $N$ agents with time synchronization errors $\delta_i, i=1,2,\dots, N$, a globally useful reachable set of the whole system under the view of agent $A_i$ based on its current local time clock $t_c^i$ is defined below:
\begin{equation}
\begin{split}
&globalReach = \bigwedge_{i=1}^N \mathcal{R}_i[t^i_{rs}, t^i_{rs} + T] \wedge \mathcal{T}, \\
&\mathcal{T} \deq (t^i_c \leq t \leq T + min\{t^i_{rs} - \delta_i - \delta_j\}, j\neq i, 1 \leq j \leq N).
\end{split}
\end{equation}
\end{definition}

For any time $t$ such that $t^i_c \leq t \leq T + min\{t^i_{rs} - \delta_i -\delta_j\}$ for $\forall~1 \leq j \leq N, i \neq j$, we have $\mathcal{R}_i(t) \subseteq \mathcal{R}_i[t^i_{rs}, t^i_{rs} + T], \forall i$. In other words, $globalReach$ contains all possible trajectories of all agents from the current local time $t^i_c$ of agent $A_i$ to the future time defined by $T + min\{t^i_{rs} - \delta_i - \delta_j\}, j\neq i, 1 \leq j \leq N$.

It should be noted that to construct a global reachable set, an agent needs to wait for all messages arrive and then decodes all these messages. This process may have an expensive computation cost especially when the number of agents increases. Since this global reachable set is only valid in an interval of time, the amount of time that is available for verify the global property may be small and not enough for the agent to perform the global safety verification. Having additional hardware for handling in parallel the processes of receiving/decoding messages is a good solution to overcome this challenge.

Using the globally useful reachable set, the global safety verification problem  is equivalent to checking whether the globally useful reachable set intersects with the global unsafe region defined by $\mathcal{U} \deq Cx \leq d$,  where $x = [x_1^T, x_2^T, \cdots , x_N^T]^T$ and $x_i$ is the state vector of agent $A_i$. The procedure for global safety verification is summarized in Algorithm \ref{arg:global-predicate-detection}.

\begin{algorithm}[]\small
    \caption{Decentralized Real-Time Global Safety Verification at Agent $A_i$.}
    \label{arg:global-predicate-detection}
		\begin{flushleft}
    \textbf{Input}: $\mathcal{U}$, ~~\% global unsafe constraints \\
    \textbf{Output}: $global\_safe, T_{global\_safe}$ ~~\% global safe flag and safe time interval in the future
		\end{flushleft}
    \begin{algorithmic}[1]
        \Procedure{Initialization}{}
            \State    $global\_safe = true$ ~~~~~\% {global safety flag}
        \EndProcedure

        \Procedure{Global Safety Verification}{}
        \If{all useful messages are available}
        \State $t^i_c = A_i.current\_time()$
        \State recheck if all messages are still useful
        \State construct globally useful reach set $globalReach$
        \If{$(globalReach \cap \mathcal{U} \neq \emptyset)$}
        \State $global\_safe = uncertain$
        \State $T_{global\_safe} = [~]$
        \Else
        \State $global\_safe = true$
        \State $T_{global\_safe} = T + min\{t_{rs}^i - \delta_i - \delta_j\}, j\neq i, 1 \leq j \leq N$
        \EndIf
        \EndIf
        \EndProcedure
    \end{algorithmic}
  \end{algorithm}

\begin{lemma}
  The decentralized real-time global safety verification algorithm is sound.
\begin{proof}
Similar to Lemma \ref{lm:collision}, the soundness of the verification algorithm is guarantee because of the soundness of the globally useful reachable set containing all possible trajectories of all agents at any time $t \in \mathcal{T}$, where $\mathcal{T} \deq (t^i_c \leq t \leq T + min\{t^i_{rs} - \delta_i - \delta_j\}, j\neq i, 1 \leq j \leq N)$.
\end{proof}
\end{lemma}

\section{Case study}
\label{sec:evaluation}

The decentralized real-time safety verification for distributed CPS proposed in this paper is implemented in Java as a package called $drreach$. This package is currently integrated as a library in StarL, which is a novel platform-independent framework for programming reliable distributed robotics applications on Android~\cite{DBLP:journals/corr/LinM15}. StarL is specifically suitable for controlling a distributed network of robots over WiFi since it provides many useful functions and sophisticated algorithms for distributed applications. In our approach, we use the reliable communication network of StarL which is assumed to be asynchronous and peer-to-peer. There may be message dropouts and transmission delays; however, every message that an agent tries to send is eventually delivered with some time guarantees. All experimental results of our approach are reproducible and available online at: \url{http://www.verivital.com/rtreach/}.

\subsection{Experiment setup}

\begin{figure}[t]
	\centering
		\includegraphics[width = 0.7\textwidth]{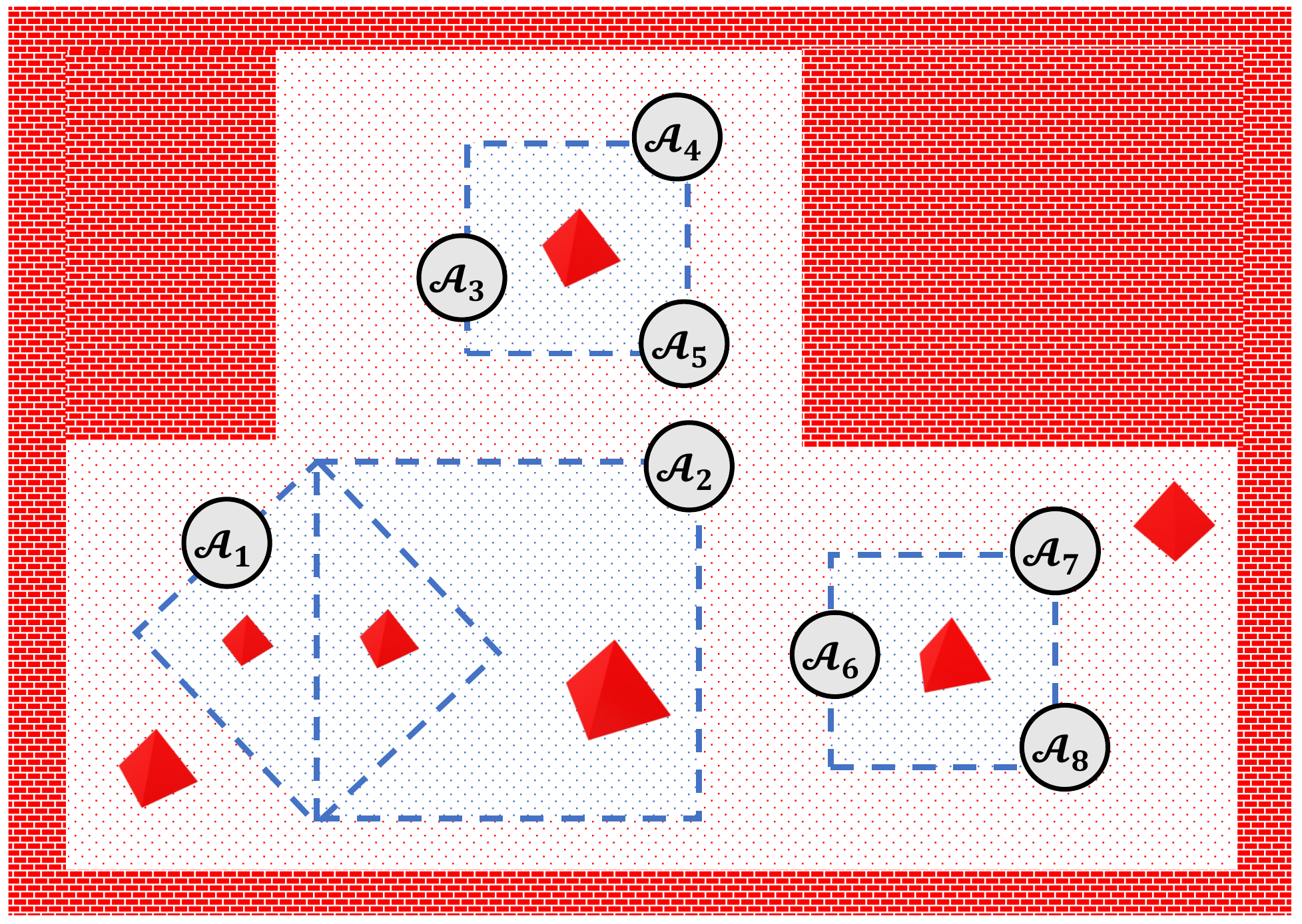}
		\caption{Distributed Search Application Using Quadcopters.}
		\vspace{-1em}
	\label{fig:case-study}
\end{figure}
We evaluate the proposed approach via a distributed search application using quadcopters\footnote{A video recording is available at: \url{https://youtu.be/YC_7BChsIf0}} in which each quadcopter executes its search mission provided by users as a list of way-points depicted in Figure \ref{fig:case-study}. These quadcopters follow the way-points to search for some specific objects. For safety reasons, they are required to work only in a specific region defined by users. In this case study, the quadcopters are controlled to operate at the same constant altitude. It has been shown from the experiments that the proposed approach is promisingly scalable as it works well for a different number of quadcopters. We choose to present in this section the experimental results for the distributed search application with eight quadcopters.

The first step in our approach is locally computing the reachable set of each quadcopter using face-lifting method. The quadcopter has nonlinear motion dynamics given in Equation \ref{model:original} in which $\theta$, $\phi$, and $\psi$ are the pitch, roll, and yaw angles, $f = \Sigma_{i=1}^4T_i$ is the sum of the propeller forces, $m$ is the mass of the quadcopter and $g = 9.81 m/s^2$ is the gravitational acceleration constant. As the quadcopter is set to operate on a constant altitude, we have $\ddot{z} = 0$ which yields the following constraint: $f = \frac{mg}{cos(\theta)cos(\phi)}$. Let $v_x$ and $v_y$ be the velocities of a quadcopter along with x- and y- axes. Using the constraint on the total force, the motion dynamics of the quadcopter can be rewritten as a $4$-dimensional nonlinear ODE as depicted in Equation \ref{model:simplified}.

\noindent\begin{minipage}{.65\linewidth}
\begin{equation}\label{model:original}
    \begin{split}
&\ddot{x} = \frac{f}{m}(sin(\psi)sin(\phi) + cos(\psi)sin(\theta)cos(\phi)), \\
&\ddot{y} = \frac{f}{m}(sin(\psi)sin(\theta)cos(\phi) - sin(\phi)cos(\psi)), \\
&\ddot{z} = \frac{f}{m}cos(\theta)cos(\phi) - g,
\end{split}
  \end{equation}
\end{minipage}%
\begin{minipage}{.35\linewidth}
  \begin{equation} \label{model:simplified}
    \begin{split}
&\dot{x} = v_x, \\
&\dot{v}_x = g tan(\theta), \\
&\dot{y} = v_y, \\
&\dot{v}_y = g \frac{tan(\phi)}{cos(\theta)}.
\end{split}
  \end{equation}
\end{minipage}

%
%

A PID controller is designed to control the quadcopter to move from its current position to desired way-points. Details about the controller parameters can be found in the available source code. The PID controller has a control period of $T_c = 200$ milliseconds. In every control period, the control inputs pitch $(\theta)$ and roll $(\phi)$ are computed based on the current positions of the quadcopter and the current target position (i.e., the current way-point it needs to go). Using the control inputs, the current positions and velocities given from GPS and the motion dynamics of the quadcopter, the real-time reachable set computation algorithm (Algorithm \ref{arg:real-time-reach}) is executed \emph{inside} the controller. This algorithm computes the reachable set of a quadcopter from its current local time to the next $T = 2$ seconds. The allowable run-time for this algorithm is $T_{runtime} = 10$ milliseconds. The local safety property is verified by the real-time reachable set computation algorithm at run-time. The computed reachable set is then encoded and sent to another quadcopter. When a reachable set message arrives, the quadcopter decodes the message to reconstruct the current reachable set of the sender. The GPS error is assumed to be $2\%$. The time-synchronization error between the quadcopters is $\delta = 3$ milliseconds.
\begin{figure*}
  \includegraphics[width=0.8\textwidth]{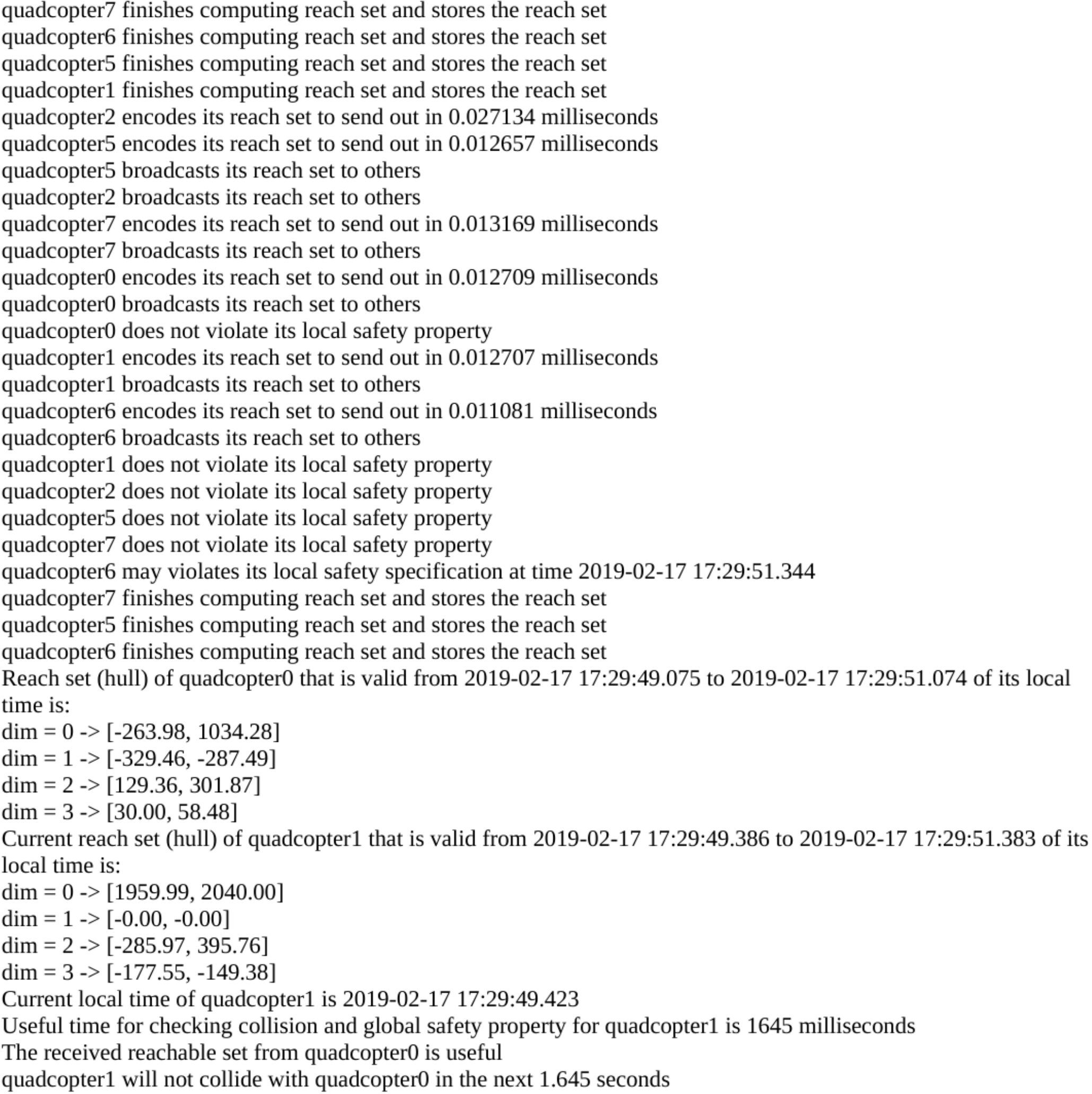}
	\vspace{-0.5em}
  \caption{A sample of events for verifying local safety property and collision occurrence.}
	\vspace{-0.25em}
  \label{fig:events}
\end{figure*}
\begin{figure}[tbp]
	\centering
		\includegraphics[width=0.8\textwidth]{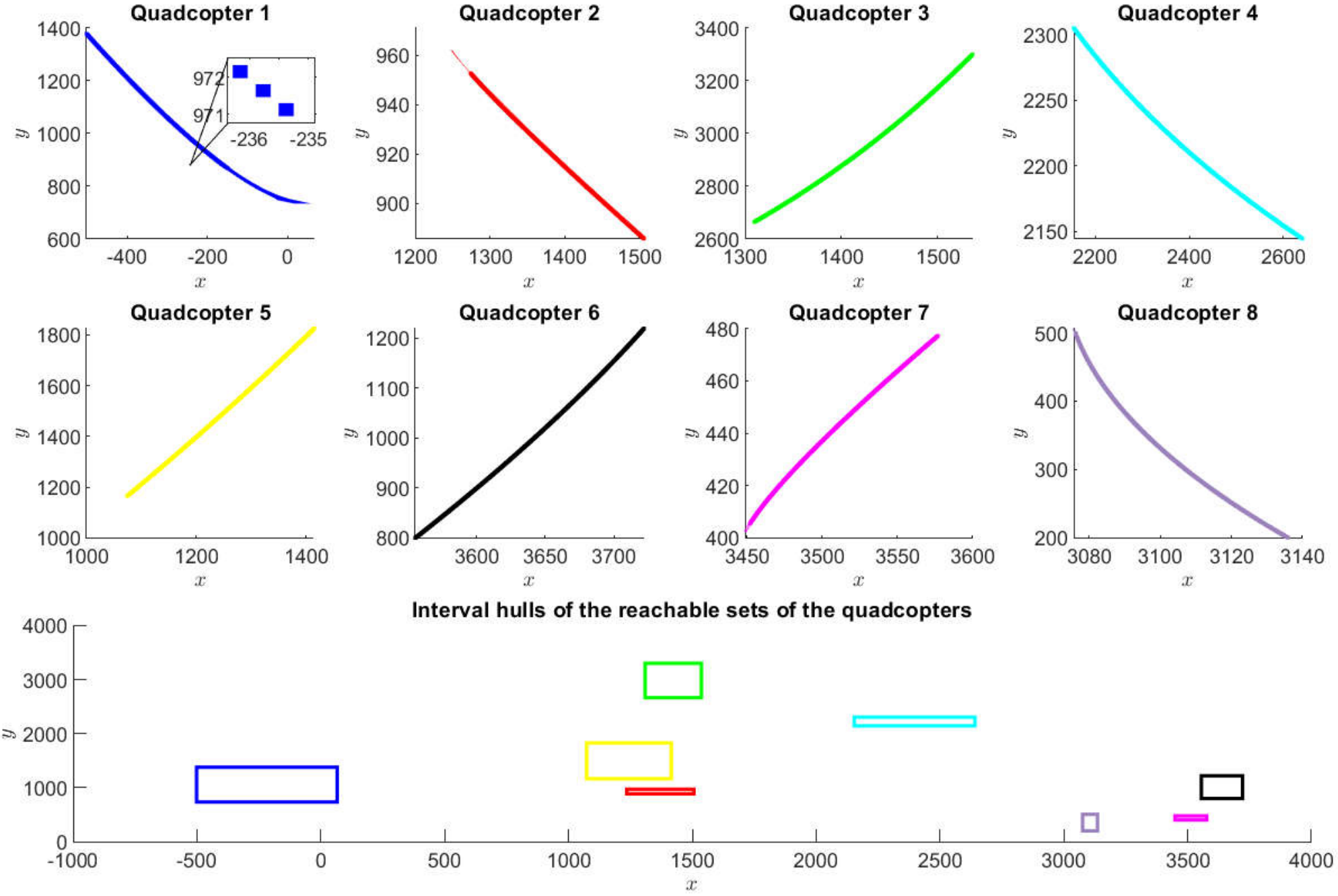}
		\caption{One sample of the reachable sets of eight quadcopters in $[0, 2s]$ time interval and their interval hulls.}
		\vspace{-1em}
	\label{fig:reach-set}
  \end{figure}
We want to verify in real-time: 1) local safety property for each quadcopter; 2) collision occurrence; and 3) geospatial free property. The local safety property is defined by $v_x \leq 500$, i.e., the maximum allowable velocities along the x-axis of two arbitrary quadcopters are not larger than $500 m/s$. The collision is checked using the minimum allowable distance between two arbitrary quadcopters $d_{min} = 100$. The geospatial free property requires that the some quadcopters never go into a specific region at the same time.
\subsection{Verifying local safety property and collision occurrence}
Figure \ref{fig:events} presents a sample of a sequence of events happening in the distributed search application. One can see that each quadcopter can determine based on its local clocks if there is no collision to some known time in the future. In addition, the local safety property can also be verified at run-time. For example, in the figure, the quadcopter $1$ receives a reachable set message from the quadcopter $0$ which is valid from $17:29:49.075$ to $17:29:51.074$ of the quadcopter $0$'s clock. After decoding this message, taking into account the time-synchronization error $\delta$, quadcopter $1$ realizes that the received reachable set message is useful for checking collision for the next $1.645$ seconds of its clock. After checking collision, quadcopter 1 knows that it will not collide with the quadcopter 0 in the next 1.645 seconds (based on its clock).

It should be noted that we can intuitively verify the collision occurrences by observing the intermediate reachable sets of all quadcopters and their interval hulls. The \emph{intermediate} reachable sets of the quadcopters in every $[0, 2s]$ time interval computed by the real-time reachable set computation algorithm (i.e., Algorithm \ref{arg:real-time-reach}) is described in Figure \ref{fig:reach-set}. The zoom plot within the figure presents a very short-time interval reachable set of the quadcopters. We note that the intermediate reachable set of a quadcopter is represented as a list of hyper-rectangles and is used for verifying the local safety property at run-time. The reachable set that is sent to another quadcopter is the interval hull of these hyper-rectangles. The intermediate reachable set cannot be transferred via a network since it is very large (i.e., hundreds of hyper-rectangles). The interval hull of all hyper-rectangles contained in the intermediate reachable set covers all possible trajectories of a quadcopter in the time interval of $[0, 2s]$. Therefore, it can be used for safety verification. One may question why we use the interval hull instead of using the convex hull of the reachable set since the former one results in a more conservative result. The reason is that we want to perform the safety verification online, convex hull of hundreds of hyper-rectangles is a time-consuming operation. Therefore, in the real-time setting, interval hull operation is a suitable solution. From the figure, we can see that the interval hulls of the reachable set of all quadcopters do not intersect with each other. Therefore, there is no collision occurrence (in the next 2 seconds of global time).
\vspace{-1em}
\begin{table}[]
  \resizebox{\textwidth}{!}{
\begin{tabular}{|l|c|c|c|c|c|c|c|c|}
\hline
\textbf{Time} & \textbf{Quad. 1} & \textbf{Quad. 2} & \textbf{Quad. 3} & \textbf{Quad. 4} & \textbf{Quad. 5} & \textbf{Quad. 6} & \textbf{Quad. 7} & \textbf{Quad. 8} \\ \hline
\textit{\textbf{Ecoding Time $\tau_{e}$ (ms)}} & 0.058 & 0.055 & 0.0553 & 0.0525 & 0.0557 & 0.0583 & 0.0584 & 0.0597 \\ \hline
\textit{\textbf{Decoding Time $\tau_{d}$ (ms)}} & 0.0169 & 0.0193 & 0.0197 & 0.019 & 0.0210 & 0.0181 & 0.0177 & 0.022 \\ \hline
\textit{\textbf{Transferring Time $\tau_{tf}$ (ms)}} & 2.64 & 2.48 & 1.42 & 1.11 & 1.12 & 1.08 & 1.05 & 1.13 \\ \hline
\textit{\textbf{Collision Checking Time $\tau_c$ (ms)}} & 0.04 & 0.05 & 0.07 & 0.05 & 0.03 & 0.07 & 0.07 & 0.14 \\ \hline
\textit{\textbf{Total Verification Time $VT$ (ms)}} & 28.9363 & 27.9 & 20.6232 & 18.3055 & 18.2527 & 18.235 & 18.0223 &19.1037 \\ \hline
\end{tabular}}
\caption{The average encoding time $\tau_e$, decoding time $\tau_d$, transferring time $\tau_{tf}$, collision checking time $\tau_c$ and total verification time $VT$ of the quadcopters.}
\label{tab:times}
\vspace{-2em}
\end{table}

Since we implement the decentralized real-time safety verification algorithm inside the quadcopter's controller, it is important to analyze whether or not the verification procedure affects the control performance of the controller. To reason about this, we measure  the average encoding, decoding, transferring and collision checking times for all quadcopters using $100$ samples which are presented in Table \ref{tab:times}. We note that the transferring time $\tau_{tf}$ is the average time for one message transferred from other quadcopters to the $i^{th}$ quadcopter. It can be seen that the encoding, decoding and collision checking times at each quadcopter constitute a tiny amount of time. The total verification time is the sum of the reachable set computation, encoding, transferring, decoding and collision checking times. Note that the allowable runtime for reachable set computation algorithm is specified by users as $T_{runtime} = 10$ milliseconds. Therefore, the (average) total time for the safety verification procedure on each quadcopter is $VT_i = T_{runtime} + \tau_e^i + (N-1)\times(\tau_{tf}^i + \tau_{d}^i + \tau_{c}^i)$, where  $i = 1, 2, \ldots, N$, and $N$ is the number of quadcopters. As shown in the Table, the (average) total verification time for each quadcopter is small ($< 30$ milliseconds), compared to the control period $T_c = 200$ milliseconds. Besides, from the experiment, we observe that the computation time for the control signal of the PID controller $\tau_{control}^i$ (not presented in the table) is also small, i.e., from $5$ to $10$ milliseconds. Since $VT_i + \tau_{control}^i < T_c/4 = 50$ milliseconds, we can conclude that the verification procedure does not affect the control performance of the controller.

Interestingly, from the verification time formula, we can estimate the range of the number of agents that the decentralized real-time verification procedure can deal with. The idea is that, in each control period $T_c$, after computing the control signal, the remaining time bandwidth $T_c - \tau_{control}$ can be used for verification. Let $\bar{\tau}_{e}(\underbar{$\tau$}_{e})$, $\bar{\tau}_{tf}(\underbar{$\tau$}_{tf})$, $\bar{\tau}_{d}(\underbar{$\tau$}_{d})$, $\bar{\tau}_{c}(\underbar{$\tau$}_{c})$ be the maximum (minimum) encoding, transferring, decoding and collision checking times on a quadcopter, $\bar{\tau}_{control} (\underbar{$\tau$}_{control})$ be the maximum (minimum) control signal computation time for each control period $T_c$, then the number of agents that the decentralized real-time safety verification procedure can deal with (with assumption that the communication network works well) satisfies the following constraint:
\begin{equation}
 \frac{T_c -  \bar{\tau}_{control} - T_{runtime} - \bar{\tau}_e}{\bar{\tau}_{tf} + \bar{\tau}_d + \bar{\tau}_c} + 1  \leq  N \leq \frac{T_c -  \underbar{$\tau$}_{control} - T_{runtime} - \underbar{$\tau$}_e}{\underbar{$\tau$}_{tf} + \underbar{$\tau$}_d + \underbar{$\tau$}_c} + 1.
\end{equation}

Let consider our case study, from the Table, we assume that $\bar{\tau}_e = 0.0597$, $\underbar{$\tau$}_e = 0.0525$, $\bar{\tau}_{tf} = 2.64$, $\underbar{$\tau$}_{tf} = 1.05$, $\bar{\tau}_d = 0.022$, $\underbar{$\tau$}_d = 0.0169$, $\bar{\tau}_c = 0.14$, $\underbar{$\tau$}_c = 0.03$ milliseconds. Also, we assume that $\bar{\tau}_{control} = 10$ and $\underbar{$\tau$}_{control} = 5$ milliseconds. We can estimate theoretically the number of quadcopters that our verification approach can deal with is $64 \leq N \leq 168$.

\subsection{Verifying geospatial free property}

\begin{figure*}
  \includegraphics[width=0.8\textwidth]{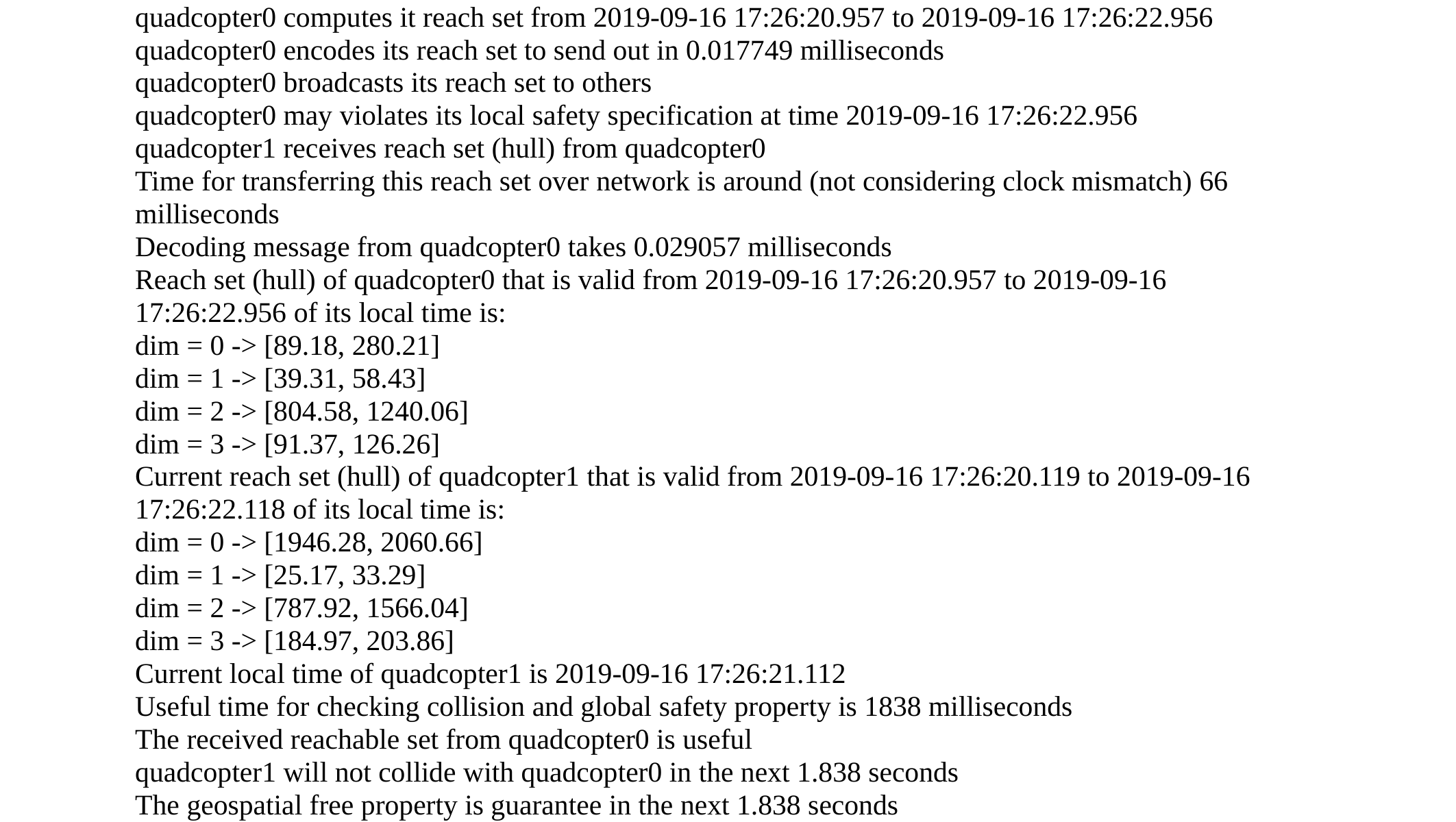}
	\vspace{-0.5em}
  \caption{A sample of events for verifying geospatial free property.}
	\vspace{-0.25em}
  \label{fig:geospatial-events}
\end{figure*}

To illustrate how our approach verifies the global behavior of a distributed CPS, we consider the geospatial free property which requires that the some (or all) quadcopters never go into a specific region at the same time. For simplification, we reconsider the distributed search application with two quadcopters (quad 1 and quad 2) whose forbidden region is defined by $900 < x_0 < 1200 \wedge 900 < x_1 < 1200$. Figure \ref{fig:geospatial-events} describes a sample of events describing that the quadcopter 2 can verify based on its local clock that it will not collide with the quadcopter 1 and the global geospatial free property is guarantee in the next $1.838$ seconds.

\section{Discussion}
\label{sec:discussion}
\begin{figure*}[t!]
  \centering
      \includegraphics[width=0.8\textwidth]{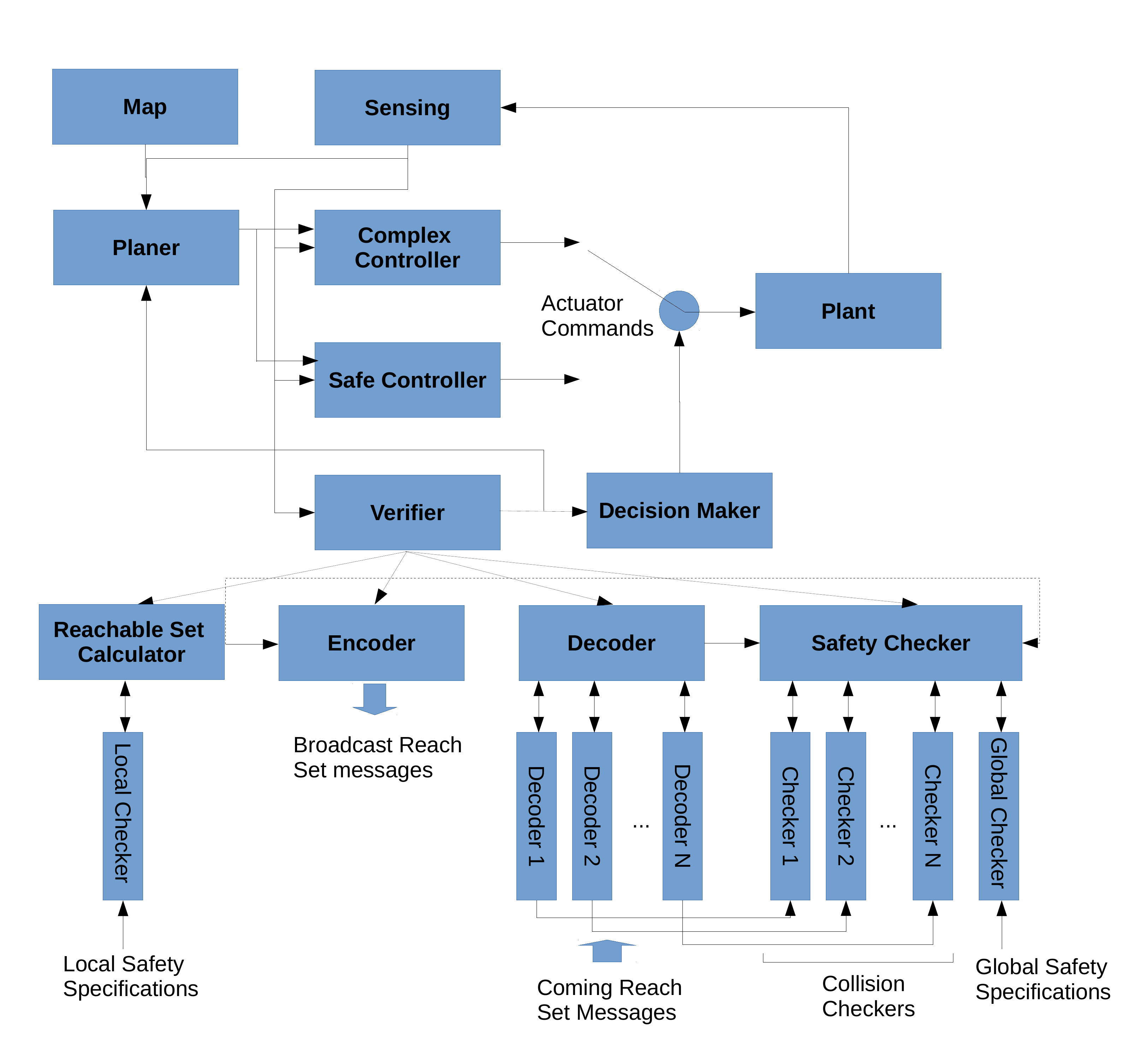}
			\vspace{-1em}
      \caption{Software architecture for deploying decentralized real-time safety verification approach on a real platform.}
  \label{fig:softwareArchitecture}
\end{figure*}

The current implementation of our approach deploys the safety verifier of each agent inside the controller, and a single thread is used to execute the control and verification tasks. The main drawback of this implementation is that it may decrease the overall performance of the controller and even cause the controller to crash. To prevent this happens, in practice, the controller and verifier should be implemented in two separate software components. In this case, the computation burden for safety checks in the verifier does not affect the performance of the controller. The control task and the verification task can be executed efficiently in parallel as depicted in Figure \ref{fig:softwareArchitecture}. More importantly, this software architecture adopts the architecture of a fault-tolerant system \cite{goodloe2010monitoring} to prevent the propagation of failure from one component to others. It also benefits the use of simplex-architecture for safety control in the case of dangerous circumstances. 

As shown in Figure \ref{fig:softwareArchitecture}, the verifier component consists of four sub-components including reachable set calculator, encoder, decoder, and safety checker. These sub-components should also be implemented conveniently for parallel execution. The local safety property is verified inside the reachable set calculator at runtime. As the number of reachable set messages needs to be decoded increases with the number of participating agents, it is necessary to have multiple decoders working in parallel. These decoders listen to upcoming reachable set messages on different ports assigned to them by the verifier and immediately decode any arrived message. This parallel decoding helps to reduce the decoding time significantly. The decoded reachable sets are then sent to the safety checker containing multiple checkers run in parallel in which each checker is responsible for checking collision between the agent with another. The $i^{th}$ checker and the $i^{th}$ decoder is a pair worker, i.e., the checker only waits for the decoded reachable set of its corresponding co-worker. Therefore, the pair to pair collision detection task can be done very quickly. The safety checker also has a global checker which is responsible for checking global properties. The global checker is only triggered when the decoder component finishes decoding all arrived reachable set messages. For this reason, having parallel working decoders is essential to speed up the overall verification time which is required to be very small to work in the real-time setting.

To analyze how fast our verification technique can achieved with the proposed software architecture, let $\bar{\tau}_{rs}, \bar{\tau}_e, \bar{\tau}_{tf}$ and $\bar{\tau}_d$ respectively be the worst case times of reachable set computation, encoding, transferring and decoding, $\bar{\tau}_{cc}$ and $\bar{\tau}_{gc}$ be the worst case times of peer-to-peer collision detection and global safety verification. For a system with $N$ agents, the total worst-case verification time is $\bar{\tau}_{total} = \bar{\tau}_{rs} + \bar{\tau}_e + \bar{\tau}_{tf} + \bar{\tau}_{d} + \bar{\tau}_{cc} + \bar{\tau}_{gc}$. If we do the verification in \emph{sequential way}, i.e., using only one port for reachable set communication and one checker for all peer-to-peer collision detection and global safety verification, the total worst-case verification is: $\bar{\tau}^*_{total} = \bar{\tau}_{rs} + \bar{\tau}_e + \bar{\tau}_{tf} + N\bar{\tau}_{d} + N\bar{\tau}_{cc} + \bar{\tau}_{gc} >> \bar{\tau}_{total}$.

\vspace{0.5em}

{\bf Scalability}. From the above discussion, one can see that the software architecture plays an important role when we implement our approach in a real platform. In practice, if each participating agent has the powerful hardware for communication and computation, and the software for our approach is implemented in a parallel manner as proposed above, then the worst-case verification time does not depend on the number of agents in the system.
Therefore, our decentralized real-time safety verification approach is scalable for systems with a large number of agents. Also, the proposed software architecture is especially useful in the case that there are losses of reachable set messages. In this hazardous situation, the agent still has some partial information to check if a collision occurs based on the available, reachable set messages. Therefore, the planner still can re-perform path planning algorithm based on the current information and past information it has to find the safest path for the agent for this incomplete information situation.

\section{Related Work}
\label{sec:related_work}
Our work is inspired by the static and dynamic analysis of timed distributed traces~\cite{duggirala2012static} and the real-time reachability analysis for verified simplex design~\cite{bak2014real}. The former one proposes a sound method of constructing a global reachable set for a distributed CPS based on the recorded traces and time synchronization errors of participating agents. Then the global reachable set is used to verify a global property using Z3~\cite{de2008z3}. This method can be considered to be a \emph{centralized analysis} where the reachable set of the whole system is constructed and verified by one analyzer. Such a verification approach is offline which is fundamentally different from our approach as we deal with online verification in a decentralized manner. Our real-time verification method borrows the face-lifting technique developed in~\cite{bak2014real} and applies it to a distributed CPS.

Another interesting aspect of real-time monitoring for linear systems was recently published in~\cite{chen2017model}. In this work, the authors proposed an approach that combines offline and online computation to decide if a given plant model has entered an uncontrollable state which is a state that no control strategy can be applied to prevent the plant go to the unsafe region. This method is useful for a single real-time CPS, but not a distributed CPS with multiple agents.

Additionally, there has been other significant works for verifying distributed CPS. Authors of \cite{eidson2012distributed, tang2012unified, zhang2008reconfigurable} presented a real-time software for distributed CPS but did not perform a safety verification of individual components and a whole system. The works presented in~\cite{johnson2012parametrized, bae2015designing, kumar2012hybrid} can be used to verify distributed CPS, but they do not consider a real-time aspect. An interesting work proposed in~\cite{loos2011adaptive} can formally model and verify a distributed car control system against several safety objectives such as collision avoidance for an arbitrary number of cars. However, it does not address the verification problem of distributed CPS in a real-time manner. The novelty of our approach is that it can over-approximate of the reachable set of each agent whose dynamics are non-linear with a high precision degree in real-time.

The most related work to our scheme was recently introduced in~\cite{liu2017provably}. The authors proposed an online verification using reachability analysis that can guarantee safe motion of mobile robots with respective to walking pedestrians modeled as hybrid systems. This work utilizes CORA toolbox~\cite{althoff2015introduction} to perform reachability analysis while our work uses a face-lifting technique. However, this work does not consider the time-elapse for encoding, transferring and decoding the reachable set messages between each agent, which play an important role in distributed systems.

\vspace{-1em}
\section{Conclusion and Future Work}
\label{sec:conclusion}

We have proposed a decentralized real-time safety verification method for distributed cyber-physical systems. By utilizing the timing information and the reachable set information from exchanged reachable set messages, a sound guarantee about the safety of the whole system is obtained for each participant based on its local time. Our method has been successfully applied for a distributed search application using quadcopters built upon StarL framework. The main benefit of our approach is that it allows participants to take advantages of formal guarantees available locally in real-time to perform intelligent actions in dangerous situations. This work is a fundamental step in dealing with real-time safe motion/path planing for distributed robots. For future work, we seek to deploy this method on a real-platform and extend it to distributed CPS with heterogeneous agents where the agents can have different motion dynamics and thus they have different control periods. In addition, the scalability of the proposed method can be improved by exploiting the benefit of parallel processing, i.e., each agent handles multiple reachable set messages and checks for collision in parallel.

\section*{Acknowledgments}
The material presented in this paper is based upon work supported by the Air Force Office of Scientific Research (AFOSR) through contract number FA9550-18-1-0122 and the Defense Advanced Research Projects Agency (DARPA) through contract number FA8750-18-C-0089. The U.S. Government is authorized to reproduce and distribute reprints for Government purposes notwithstanding any copyright notation thereon. The views and conclusions contained herein are those of the authors and should not be interpreted as necessarily representing the official policies or endorsements, either expressed or implied, of AFOSR or DARPA.


\normalsize
\let\oldbibliography\thebibliography
\renewcommand{\thebibliography}[1]{\oldbibliography{#1}
\setlength{\itemsep}{0pt}} 
\bibliographystyle{splncs03}
\bibliography{tran}  




%






\end{document}